\documentstyle[epsfig]{cupconf}

\ifoldfss
\else
  \ifnfssone
    \newmathalphabet{\mathit}
      \addtoversion{normal}{\mathit}{cmr}{m}{it}
      \addtoversion{bold}{\mathit}{cmr}{bx}{it}
    \newmathalphabet{\mathcal}
      \addtoversion{normal}{\mathcal}{cmsy}{m}{n}
    \else
    \ifnfsstwo
    \fi
  \fi
\fi

\def\hexnumber#1{\ifcase#1 0\or1\or2\or3\or4\or5\or6\or7\or8\or9\or
 A\or B\or C\or D\or E\or F\fi }

\makeatletter
\ifx\CUP@mtlplain@loaded\undefined
\else
\fi
\makeatother
 \makeatletter
 \ifx\CUP@mtlplain@loaded\undefined
   \font\tenbmi=cmmib10 at 10pt
   \font\sevenbmi=cmmib10 at 7pt
   \font\fivebmi=cmmib10 at 5pt

   \newfam\bmifam
   \textfont\bmifam=\tenbmi
   \scriptfont\bmifam=\sevenbmi
   \scriptscriptfont\bmifam=\fivebmi
   
 \fi
 \makeatother
\ifnfsstwo

\fi
\ifnfssone

\fi
\ifoldfss

\fi

\mathchardef\varLambda="0103

\makeatletter
\ifx\CUP@mtlplain@loaded\undefined
\else
\fi
\makeatother
\makeatletter
\ifx\CUP@mtlplain@loaded\undefined
  \font\tenbms=cmbsy10
  \font\sevenbms=cmbsy10 at 7pt
  \font\fivebms=cmbsy10 at 5pt
  \newfam\bmsfam
  \textfont\bmsfam=\tenbms
  \scriptfont\bmsfam=\sevenbms
  \scriptscriptfont\bmsfam=\fivebms

  \edef\bsy@{\hexnumber\bmsfam}
  \mathchardef\bnabla="0\bsy@72
\fi
\makeatother
\title[Internet Services for Professional Astronomy]{Internet Services for Professional Astronomy}

\author[H. Andernach]%
{H\ls E\ls I\ls N\ls Z\ns A\ls N\ls D\ls E\ls R\ls N\ls A\ls C\ls H$^1$}

\affiliation{$^1$Depto.~de Astronom\'\i a, IFUG, Universidad de Guanajuato,
Guanajuato, C.P. 36000, Mexico \\
Email:~~{\tt heinz@astro.ugto.mx} }

\def\deg{$^{\circ}$}
\def\ltsim{\raise 2pt \hbox {$<$} \kern-1.1em \lower 4pt \hbox {$\sim$}}
\def\gtsim{\raise 2pt \hbox {$>$} \kern-1.1em \lower 4pt \hbox {$\sim$}}

\def\tts{\small\tt }

\begin{document}
\ifnfssone
\else
  \ifnfsstwo
  \else
    \ifoldfss
      \let\mathcal\cal
      \let\mathrm\rm
      \let\mathsf\sf
    \fi
  \fi
\fi

\hyphenation{SIMBAD}

\maketitle

\vspace*{-5.3cm}
\begin{center}
\begin{footnotesize}\baselineskip 10 pt\noindent
To appear in ``Astrophysics with Large Databases in the Internet Age'' \\
Proc.~~{\it IX$^{th}$ Canary Islands Winter School on Astrophysics}, Tenerife, Spain,~ Nov.\,17--28, 1997 \\
eds.~M.~Kidger,~I.~P\'erez-Fournon,~\& F.~S\'anchez,
Cambridge University Press, 1998
\end{footnotesize}
\end{center}
\vspace*{4.1cm}

\begin{abstract}
A (subjective) overview of Internet resources relevant to
professional astronomers is given.  Special emphasis is put
on databases of astronomical objects and servers providing
general information, e.g. on astronomical catalogues,
finding charts from sky surveys, bibliographies, directories,
browsers through multi-wavelength observational archives, etc.~
Archives of specific observational data will be
discussed in more detail in other chapters of this book,
dealing with the corresponding part of the electromagnetic spectrum.
About 200 different links are mentioned, and every attempt was made
to make this report as up-to-date as possible.
As the field is rapidly growing with improved network technology,
it will be just a snapshot of the present situation.
\end{abstract}

\firstsection 

\section{Introduction}\label{intro}

During the five or so years since the advent of the {\it World Wide Web} (WWW)
the number of servers offering information for astronomers has grown
as explosively as that of other servers (cf. \cite{adorf95}).
Perhaps even more than other media, the Internet
is flooding us with an excessive amount of
information and it has become a challenge to distinguish signal from
noise. This report is yet another attempt to do this.

A web address is usually referred to as ``Universal Resource Locator'' (URL),
and starts with the characters ``{\tt http://}''.
For better readability I omit these characters here, since one of the
most common web browsers ({\tt netscape}) assumes these by default anyway,
unless other strings like ``{\tt ftp://}'' are specified.
All URL links in this overview are given in bold font. It
should be emphasized that, owing to the nature of the WWW, some of
these URLs may change without notice. The ``File Transfer Protocol'' (FTP)
will appear as {\tt ftp} in this text to coincide with the
corresponding Unix command.

A useful introduction to the basics of Internet, explaining electronic mail,
telnet, {\tt ftp}, bulletin boards, ``Netiquette'', Archie, Gopher, Veronica,
WAIS can be found e.g. in \cite{groth95} or \cite{thomas97}.
I shall not repeat these basics here, but rather
concentrate on practical tools to obtain astronomical
information from the Internet. Many search engines have been developed
for browsing WWW servers by keywords, e.g. \linebreak[4] AltaVista, Lycos, Savvy, Yahoo,
WebGlimpse ({\tts www.altavista.digital.com}, \linebreak[4]
{\tts www.lycos.com},~
{\tts \verb*Cwww.cs.colostate.edu/~dreiling/smartform.htmlC},~
{\tts www.yahoo.com}, \linebreak[4]
{\tts donkey.CS.Arizona.EDU/webglimpse}, resp.).

\begin{figure*}[!hb]
\epsfig{file=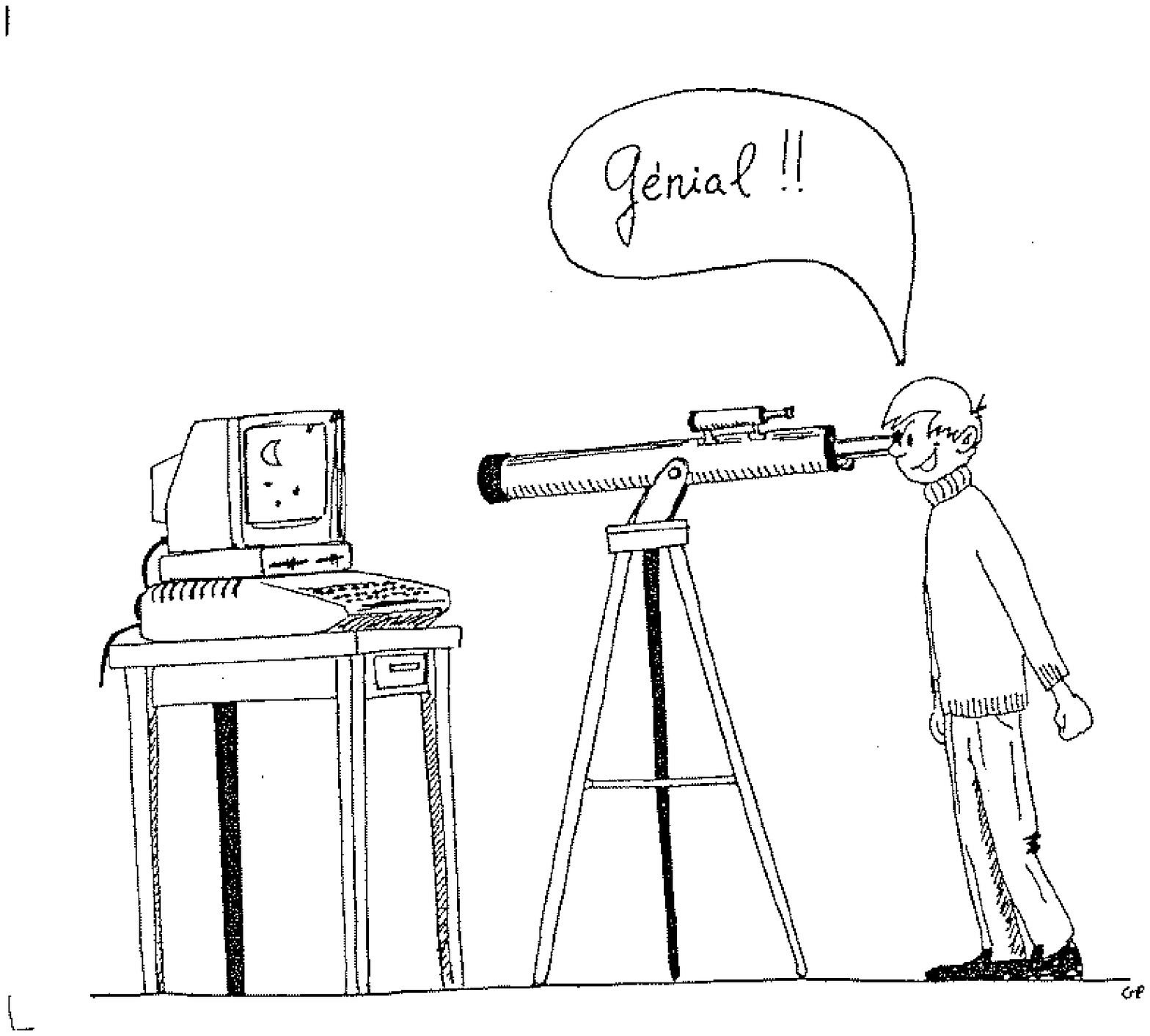,width=14.cm,bbllx=83pt,bblly=185pt,bburx=549pt,bbury=587pt,clip=,angle=-1}
\caption{A typical example of an astronomer in the Internet age.
  (Drawing~~courtesy~~Georges~Paturel)} \label{paturelfig}
\end{figure*}

A comprehensive list of URLs for ``resource discovery'' can be found
in \cite{adorf95}, at {\tts searchenginewatch.com}, or see 
{\tts www.cnet.com/Content/Features/Dlife/Search/ss06.html}
for some useful tips to define your search adequately.
For astronomical topics the flood of URLs returned is smaller at 
AstroWeb ({\tts fits.cv.nrao.edu/www/astronomy.html}) or its mirrors
({\tts cdsweb.u-strasbg.fr/astroweb.html},~
{\tts www.stsci.edu/astroweb/astronomy.html},~ \linebreak[4]
{\tts msowww.anu.edu.au/anton/astronomy.html},~
{\tts www.vilspa.esa.es/astroweb/astronomy.html}). \\
AstroWeb is a consortium of several astronomers who have been
collecting astronomy-relevant links since 1995.
However, AstroWeb does not actively skim the web for relevant
sites regularly, although it has been done occasionally. It relies mainly
on forms sent in by the authors or webmasters of those sites, and
currently collects about 2500 links.

Recent reviews related to Internet resources in astronomy were given
by Andernach, Hanisch \& Murtagh (1994), 
\cite{waw95, lisa2}, Egret \& \linebreak[4] Albrecht (1995), \cite{dissem97}, 
\cite{epub97}, and \cite{lisa3}. \linebreak[4]
Equally useful are the
proceedings of the meetings on {\it Astronomical Data Analysis
Software and Systems, I--VII}, held annually since 1991 and published
as ASP Conference Series vols. 25, 52, 61, 77, 101, 125, and 145.
For information and electronic proceedings on meetings III--VII, look at:

\smallskip

\begin{itemize}
\item {\tts cadcwww.dao.nrc.ca/ADASS/adass\_proc/adass3/adasstoc/adasstoc.html}
\item {\tts www.stsci.edu/stsci/meetings/adassIV/toc.html}
\item {\tts iraf.noao.edu/ADASS/adass.html}
\item {\tts www.stsci.edu/meetings/adassVI}
\item {\tts www.stsci.edu/stsci/meetings/adassVII}
\end{itemize}
 
\smallskip

I do not discuss services explicitly dedicated to amateurs,
although there is no well-defined boundary between professional and amateur
astronomy, and the latter can be of vital use to professionals, e.g.\ in
the field of variable stars, comets or special solar system events.
Indeed, stellar photometry --
a field traditionally dominated by amateur observers -- boasts a
database of variable star observations ({\tts www.aavso.org/internatl\_db.html})
that is the envy of many professionals
and a shining example to the professional world of \linebreak[4] cooperation,
organization and service.  A further proof of a fruitful interaction 
between professionals and amateurs is ``The Amateur Sky Survey'' (TASS) 
which plans to monitor the sky down to 14-16 mag and study variable 
stars, asteroids and comets ({\tts www.tass-survey.org}; \cite{tass98}).

A cautionary note: by its very nature of describing ``sites'' on the ``web'',
this work is much like a tourist guide with all its imperfections; hotels
or restaurants may have closed or changed their chefs, new
roads may have been opened, and beaches may have deteriorated or improved.
As similar things
happen constantly with web pages or URLs, take this work as a suggestion
only. You will be the one to adapt it to your own needs, and maintain it
as your own reference.

Although I ``visited'' virtually all links quoted in the present report,
be prepared to find obsolete, incomplete or no information at all at some
URLs. However, after having convinced yourself of missing, obsolete, or
incorrect information offered at a given site, do not hesitate to
contact the ``webmaster'' or manager of that site and make constructive
suggestions for improvement (rather than merely complain).
This is even more important if you
find a real error in database services used by a wide community. Vice versa,
in your own efforts to provide your web pages, try to avoid links to
other documents which do not exist or merely claim to be ``under construction'',
just imagine the time an interested user with a slow connection
may waste in calling such a link.

\section{Data Centres}

The two largest, general purpose data centres for astronomy world-wide
are the ``Astronomical Data Center'' (ADC; {\tts adc.gsfc.nasa.gov/adc.html}) of
NASA's Astrophysics Data Facility (ADF), and the
``Centre de Donn\'ees astronomiques de Strasbourg'' (CDS;
{\tts cdsweb.u-strasbg.fr}). They were the first institutions to
systematically collect machine-readable versions of astronomical catalogues
but have now widened their scope considerably. Several of their services
are so diverse that they will be mentioned in different sections
of this paper.
At other, medium-sized, regional data centres (e.g. in
Moscow, Tokyo, Beijing, Pune, etc.) some of the services of the two
major centres are mirrored to reduce the network load on ADC and CDS.

The increasing number of space missions has led to the creation of
mission-oriented data centres like:

\smallskip
 
\begin{itemize}
\item IPAC ({\tts www.ipac.caltech.edu})
\item STScI ({\tts www.stsci.edu})
\item ST-ECF ({\tts ecf.hq.eso.org})
\item MPE ({\tts www.rosat.mpe-garching.mpg.de})
\item ESTEC ({\tts astro.estec.esa.nl})
\item VILSPA ({\tts www.vilspa.esa.es})
\item CADC ({\tts cadcwww.dao.nrc.ca})
\end{itemize}
 
\smallskip
 
\noindent and many others.  Several major ground-based observatories
have evolved data centres offering access to their archives
of past observations. Examples are the European Southern Observatory 
({\tts archive.eso.org}), the Royal Greenwich Observatory Isaac Newton
Group of Telescopes on La Palma ({\tts archive.ast.cam.ac.uk/ingarch}),
the National Radio Astronomy Observatory (NRAO) Very Large Array
({\tts info.aoc.nrao.edu/doc/vladb/VLADB.html}),
the Westerbork Synthesis Radio Telescope (WSRT; {\tts www.nfra.nl/scissor}),
etc.

Many bulky data sets, like e.g.\ the IRAS data products, the Hubble Space
Telescope (HST) Guide Star Catalogue (GSC), etc.,
have been widely distributed on CD-ROMs. For a
comprehensive, though not complete, list of CD-ROMs in astronomy see
the ``Mediatheque'' maintained at CDS
({\tts \verb*Ccdsweb.u-strasbg.fr/mediatheque.htmC}) and D.~Golombek's
contribution to these proceedings.  

\section{Astronomical Catalogues} \label{catal}

We consider here ``astronomical catalogues'' as static, final compilations of
data for a given set of cosmic objects. According to \cite{jas89}
they can be further subclassified into
(1) observational catalogues, (2) compilation catalogues, and
(3) critical compilation catalogues and bibliographic compilation catalogues.
In class (1) we also include tables of observational data
commonly published in research papers.

The CDS maintains the most complete set of astronomical catalogues
in a publicly accessible archive at {\tts cdsweb.u-strasbg.fr/Cats.html}.
Currently $\sim$2700 catalogues and published data tables are stored,
of which $\sim$2200 are available for downloading via {\tt ftp}, and a subset
of $\sim$1760 of them are searchable through the {\tt VizieR} browser (see below).
Catalogues can be located by author name or keyword.
Since 1993 tables published in {\it Astronomy \& Astrophysics} and its
{\it Supplement Series} are stored and documented in a standard way at CDS.
As the authors are only recommended, but not obliged, to deposit their
tables at CDS, there is a small incompleteness even for recently
published tabular material. CDS is also making serious efforts in
completing its archive by converting older or missing published tables
into electronic form using a scanner and ``Optical Character Recognition''
(OCR) software. However, catalogues prepared in such a way (as stated
in the accompanying documentation file) should be treated
with some caution, since OCR is never really free of errors and careful
proof-reading is necessary to confirm its conformity with the original.

Access to most catalogues is offered via anonymous {\tt ftp}
to {\tts cdsarc.u-strasbg.fr} in the subdirectory {\tts /pub/cats}. This
directory is further subdivided into nine sections of catalogues
(from stellar to high-energy data), and the ``J'' directory for smaller
tables from journals.  These subdirectories are directly named
after their published location, e.g., {\tts /pub/cats/J/A+AS/90/327}  has
tabular data published in A\&AS Vol. 90, p.\ 327.

There are other useful commands on the CDS node {\tts simbad} which are also
available without a ``proper'' SIMBAD account (\S\ref{objbas}).
You can telnet to
{\tts simbad.u-strasbg.fr}, login as {\tt info}, give {\tt <CR>} as password.
This account allows one to query the ``Dictionary of Nomenclature of Celestial
Objects'' and provides comments on the inclusion (or not) in SIMBAD of objects
with a certain acronym. Other useful commands are e.g.
{\tt findcat}, allowing one to locate electronic catalogues by author or keyword,
{\tt findacro} to resolve acronyms of object designations (\S\ref{Dic}),
{\tt findgsc} to search the Guide Star Catalogue (GSC~1.1), and
{\tt findpmm}
to search in the USNO-A1.0 catalogue of $\sim$5 10$^8$ objects
(\S\ref{ocatdss}).
The commands {\tt simbib} and {\tt simref} are useful to interrogate
the SIMBAD bibliography remotely (by author's names or words in paper titles)
or resolve the 19-digit ``refcodes'' (\S\ref{bibserv}).
The syntax of all commands can be checked by typing~ {\tts command -help}.
Users with frequent need for these utilities may install these commands
on their own machine, by retrieving the file
{\tts cdsarc.u-strasbg.fr/pub/cats/cdsclient.tar.Z} ($\sim$40 kb).
This allows them
to access the above information instantaneously from the command line.

NASA-ADC and CDS maintain mirror copies of their catalogue collections
(see the URL {\tts adc.gsfc.nasa.gov/adc/adc\_archive\_access.html}).
A major fraction of the CDS and NASA-ADC catalogue collection is also
available at the Japanese ``Astronomical Data Analysis Center'' (ADAC;
{\tts adac.mtk.nao.ac.jp}), at the Indian ``Inter-University Centre for
Astronomy \& Astrophysics'' (IUCAA; {\tts www.iucaa.ernet.in/adc.shtml},
click on ``Facilities''), and at the Beijing Astronomical Data
Center ({\tts www.bao.ac.cn/bdc}) in China.

NASA-ADC has issued a series of four CD-ROMs with collections of
the ``most popular'' (i.e. most frequently requested) catalogues in
their {\tt ftp} archive. Some CDs
(see {\tts adc.gsfc.nasa.gov/adc/adc\_other\_media.html}) come with a
simple software to browse the catalogues. However, the
CD-ROM versions of these catalogues are static, and errors found
in them (see errata at {\tts ftp://adc.gsfc.nasa.gov/pub/adc/errors}),
are corrected only in the {\tt ftp} archive.

Many tables or catalogues published in journals of the
American Astronomical Society (AAS), like the {\it Astrophysical Journal} (ApJ),
its {\it Supplement} (ApJS), the {\it Astronomical Journal} (AJ), and also the
{\it Publications of the Astronomical Society of the Pacific} (PASP), have
no longer appeared in print since 1994. Instead, the articles often present
a few sample lines or pages only, and the reader is referred to the AAS
CD-ROM to retrieve the full data.
As these CD-ROMs are published only twice a year, in parallel
with the journal subscription, this would imply that the reader has to wait up
to six months or more to be able to see the data. At the time of
writing neither the printed papers nor the electronic ApJ and AJ mention
that the CD-ROM data are also available via anonymous {\tt ftp} (at
{\tts ftp://ftp.aas.org/cdrom}) before or at the time of the
publication on paper. The {\tt ftp} service appears much more practical
than the physical CD-ROMs for active researchers (i.e. probably the
majority of readers), and perhaps for this reason AAS has decided
that CD-ROMs will not be issued beyond Vol.~9 (Dec. 1997;
see also {\tts www.journals.uchicago.edu/AAS/cdrom}). It would be
desirable if readers of electronic AAS journals were able to access the newly
published tabular data in their entirety and more directly in the future.
With the trend to publish tabular data only in electronic form,
there is also hope that
authors may be released from the task of marking up large tables in
\TeX\ or \LaTeX\ , since the main use of these tables will be their integration
in larger databases, processing with \TeX\ -incompatible programming
languages, or browsing on the user's computer screen, where \TeX\ formatting
symbols would only be disturbing \linebreak[4] (cf.\ also \S\ref{future}).

Most of the AAS CD-ROM data are also ingested in the catalogue archives
of CDS and ADC. However, many other catalogues, not necessarily available
from these two centres, can be found within various other services,
e.g. in CATS, DIRA2, EINLINE, HEASARC, STARCAT (see \S\ref{pseudo}),
or the LANL/SISSA server (\S\ref{ppserv}), etc..
In particular, the present author has spent much
effort to collect (or recover via OCR) many datasets (now almost 800)
published as tables in journals (see
{\tts \verb*Ccats.sao.ru/~cats/doc/Andernach.htmlC} and
{\tts \verb*Ccats.sao.ru/~cats/doc/Ander_noR.htmlC}).
About half of these were kindly provided upon request by the the authors of
the tables.
The others were prepared either via scanning and OCR by the present author,
applying careful proof-reading, or they were found on the LANL/SISSA
preprint server and converted from \TeX\ (or even PostScript) to ASCII
format by the present author. There is no master database
that would indicate on which of these servers a certain catalogue is available.

\section{Retrieving Information on Objects} \label{objretr}

We may distinguish several facets to the task of retrieving information on
a given astronomical object. To find out what has been published on that
object, a good start is made by consulting the various object
databases (\S\ref{objbas}) and looking at the bibliographical
references returned by them.  Some basic data on the object will also
be offered, but to retrieve further published measurements a detailed search in
appropriate catalogue collections will be necessary (\S\ref{pseudo}).
Eventually one may be interested in retrieving and working with
raw or reduced data on that object such as spectra, images, time series,
etc., which may only be found in archives of
ground- or space-based observatories (\S\ref{archives}).

Only objects outside the solar system will be discussed in this
section. Good sites for information on solar system objects are
NASA's ``Planetary Data System'' (PDS) at JPL (\cite{pds96};
{\tts pds.jpl.nasa.gov} and {\tts ssd.jpl.nasa.gov}), the
``Lunar and Planetary Institute''
(LPI; {\tts cass.jsc.nasa.gov/lpi.html}), the ``Solar Physics'' and
``Planetary Sciences'' links at the NSSDC
({\tts nssdc.gsfc.nasa.gov/solar} and {\tts .../planetary}), and
the ``Minor Planet Center'' ({\tts cfa-www.harvard.edu/cfa/ps/mpc.html}).
The ``Students for the Exploration and Development of Space'' homepage
also offers useful links on solar system
information ({\tts seds.lpl.arizona.edu}).

\subsection{Object Databases} \label{objbas}

Object databases are understood here as those which gather both bibliographical
references and measured quantities on Galactic and/or extragalactic
objects. There are three prime ones: SIMBAD, NED and LEDA;
the latter two are limited to extragalactic objects only.
A comparison of their extragalactic content has been given by \cite{and95a}.
All three involve
an astronomical ``object-name resolver'', which accepts and returns
identifiers; it also permits retrieval of all objects within a stated
radius around coordinates in various different systems or equinoxes.
For large lists of objects the databases can also support batch jobs,
which are prepared according to specific formats and submitted via
email. The results can either be mailed back
or be retrieved by the user via anonymous {\tt ftp}. While SIMBAD and NED allow
some limited choice of output format, LEDA is the only
one that delivers the result in well-aligned tables with one object per line.

SIMBAD (Set of Identifications, Measurements, and Bibliography for Astronomical
Data) is a database of astronomical objects outside the solar system, produced
and maintained by CDS. Presently SIMBAD contains 1.54 million objects under
4.4 million identifying names, cross-indexed to over 2200 catalogues.
It provides links to 95,700 different bibliographical references,
collected for stars systematically since 1950.
Presently over 90 journals are perused for SIMBAD, which is the
most complete database for Galactic objects
(stars, HII regions, planetary nebulae, etc.), but
since 1983 it has included galaxies and other extragalactic objects
as well.
SIMBAD is not quite self-explanatory; its user's guide
can be retrieved from
{\tts ftp://cdsarc.u-strasbg.fr/pub/simbad/guide13.ps.gz}, \linebreak[4]
or may be consulted interactively at
{\tts simbad.u-strasbg.fr/guide/guide.html}.

Access to SIMBAD requires a password, and applications may be sent by email
to \linebreak[4] {\tts question@simbad.u-strasbg.fr}.
By special agreement, access is free for astronomers affiliated
to institutions in Europe, USA and Japan, while users from
other countries are charged for access. The {\tt telnet} address of
SIMBAD is {\tts simbad.u-strasbg.fr} and its web address in Europe is
{\tts simbad.u-strasbg.fr/Simbad}.
It has a mirror site in USA, at {\tts simbad.harvard.edu}.

The ``NASA/IPAC Extragalactic Database'' (NED, \cite{ned95}) currently
contains positions,
basic data, and over 1,275,000 names for 767,000 extragalactic objects,
nearly 880,000 bibliographic references to 33,000 published papers,
and 37,000 notes from catalogues and other publications, as well as
over 1,200,000 photometric measurements, and 500,000
position measurements.
NED includes 15,500 abstracts of articles of extragalactic interest
that have appeared in A\&A, AJ, ApJ, \linebreak[0] MNRAS, and PASP
since 1988, and from several other journals in more recent years.
Although NED is far more complete in extragalactic objects
than is SIMBAD, it is definitely worthwhile consulting SIMBAD
to cover the extragalactic literature for the five years in its
archives before NED commenced in 1988.
Samples of objects may be extracted from NED through filters set
by parameters like position in the sky, redshift, object type, and many others.
NED has started to provide digital images, including finding charts from
the ``Digitized Sky Survey'' (DSS) for some 120,000 of their objects.
A unique feature of NED is that the photometric data for a given object
may be displayed in a plot of the ``Spectral Energy Distribution'' (SED).
NED is accessible without charge at {\tts nedwww.ipac.caltech.edu}
or via telnet to {\tts ned.ipac.caltech.edu} (login as {\tts ned}).

The ``Lyon--Meudon Extragalactic Database'' (LEDA), created in 1983 and
maintained at Lyon Observatory, offers free access to the main
(up to 66)  astrophysical parameters for about 165,000 galaxies in the
``nearby'' Universe (i.e. typically z$<$0.3).
All raw data as compiled from literature are available, from which
mean homogenized parameters are calculated according to reduction procedures
refined every year (\cite{patu97}).
Finding charts of galaxies, at almost any scale, with or without stars from
the GSC, can be created and $\sim$74,000 images of part of these galaxies can be
obtained in PS format.
These images were taken with a video camera from the POSS-I for
identification purposes only and are of lower quality than those from the
Digitized Sky Survey (\S\ref{dssetc}). However, they have been
used successfully by the LEDA team to improve positions and shape parameters of
the catalogued galaxies. LEDA also incorporates the galaxies (20,000 up to now)
which are being detected in the ongoing ``Deep Near-IR Survey of the
Southern Sky'' (DENIS; {\tts www.strw.leidenuniv.nl/denis} or
{\tts denisexg.obspm.fr/denis/denis.html}).
A flexible query language allows the user to define and
extract galaxy samples by complex criteria. LEDA can be accessed at
{\tts www-obs.univ-lyon1.fr/leda} or via telnet to {\tts lmc.univ-lyon1.fr}
(login as {\tts leda}). The homogenized part of LEDA's data, together with
simple interrogation software, is being released on ``PGCROMs''s every
four years (1992, 1996, 2000...).

\subsection{Pseudo-Databases: Searchable Collections of Catalogues} \label{pseudo}

It should be kept in mind that object databases like SIMBAD, NED and LEDA,
generally do not include the full information contained in the CDS/ADC
collections of catalogues and tables. This is especially true for
older tables which may have become available in electronic form only
recently. The catalogues and tables frequently contain data columns not
(yet) included in the object databases. Thus the table collections should
be considered as a valuable complement to the databases. Different sites
support different levels of search of those collections.

Probably the largest number of individual catalogues ($\sim$1560)
that can be browsed from one site is that offered by {\tt VizieR} at CDS
({\tts vizier.u-strasbg.fr}).  You may select the catalogues by type of
object, wavelength range, name of space mission, etc.  An advantage
of {\tt VizieR} is that the result comes with hyperlinks (if available)
to SIMBAD or other
relevant databases, allowing more detailed inquiries on the retrieved objects.
A drawback is
that a search on many of them at the same time requires selecting
them individually by clicking on a button. An interface allowing
searches through many or all of them is under construction.

The DIRA2 service (``Distributed Information Retrieval from Astronomical
files'') at {\tts www.ira.bo.cnr.it/dira/gb}
is maintained by the ASTRONET Database Working Group in Bologna, Italy.
It provides access to data from astronomical catalogues (see the
manual at {\tts www.pd.astro.it/prova/prova.html}).
The DIRA2 database contains about 270 original catalogues of Galactic and
extragalactic data written in a DIRA-specific ASCII format.
The output of the searches are ASCII or FITS files that can be used in other
application programs.  DIRA2 allows one to plot objects in an area of
sky taken from various catalogues onto the screen with various symbols of
the user's choice. Sorting as well as selecting and
cross-identification of objects from different catalogues is possible,
but there is no easy way to search through many catalogues at a time.
The software is publicly available for various platforms.

The ``CATalogs supporting System'' (CATS; {\tts cats.sao.ru}) has been
developed at the Special Astrophysical Observatory (SAO) in Russia.
Apart from dozens of the larger general astronomical catalogues it offers
the largest collection of radio source catalogues searchable with a single
command (see chapter on Radio Astronomy by H. Andernach in these
proceedings). A search through the entire catalogue collection ``in one
shot'' is straightforward.

A set of about 100 catalogues, dominated by X-ray source
catalogues and mission logs, can be browsed at the ``High Energy
Astrophysics Science Archive Research Center'' (HEASARC) at~
{\tts heasarc.gsfc.nasa.gov/W3Browse}. The same collection is
available at the ``Leicester Database and Archive Service''
(LEDAS; {\tts ledas-www.star.le.ac.uk}).

A similar service, offered via telnet and without a web interface,
is the ``Einstein On-line Service'' (EOLS, or EINLINE) at the
Harvard-Smithsonian Center for Astrophysics (CfA). It was designed
to manage X-ray data from the {\it EINSTEIN} satellite, but it also
served in 1993/94 as a testbed for the integration of radio source catalogues.
Although EOLS is still operational with altogether 157 searchable catalogues
and observing logs, lack of funding since 1995 has prevented any
improvement of the software and interface or the integration of new catalogues.

NASA's ``Astrophysics Data System'' (ADS) offers a ``Catalog Service'' at \linebreak[4]
{\tts adscat.harvard.edu/catalog\_service.html}.
With the exception of the Minnesota Plate Scanning project
(APS, cf. \S\ref{dssimag}), all available 130 catalogues are stored
at the Smithsonian Astrophysical Observatory (SAO).
For a complete list, request ``catalogues by name'' from the catalogue service.
Not all available catalogues can be searched simultaneously.
The service has not been updated for several years and will eventually
be merged with {\tt VizieR} at CDS.

A growing number of catalogues is available in {\it dat}OZ (see
{\tts 146.83.9.18/datoz\_t.html}) at the  University of Chile as described
in \cite{ortiz98}. It offers visualization and cross-correlation tools.
The STARCAT interface at ESO ({\tts arch-http.hq.eso.org/starcat.html}) with
only 65 astronomical catalogues is still available, but has become obsolete. \linebreak[4]
ASTROCAT at CADC ({\tts cadcwww.dao.nrc.ca/astrocat}) offers
about 14 catalogues.

\subsection{Archives of Observational Data} \label{archives}

As these will be discussed in more detail in the various chapters of this
book dedicated to Internet resources in different parts of the
electromagnetic spectrum, I list only a few URLs from which the user
may start to dig for data of his/her interest.

Abstracts of sections of the book by \cite{egralb95} are available on-line at \\
{\tts cdsweb.u-strasbg.fr/data-online.html} and provide
links to several archives.
The AstroWeb consortium offers a list of currently 129 records for
``Data and Archive Centers'' at {\tts www.cv.nrao.edu/fits/www/yp\_center.html}.
STScI ({\tts archive.stsci.edu}) has been designated by NASA as a multi-mission
archive centre, focussing on optical and UV mission data sets.
It now plays a role analogous to HEASARC for high-energy data (see below)
and IPAC ({\tts www.ipac.caltech.edu}) for infra-red data. The latter
are associated in the ``Astrophysics Data Centers Coordinating Council''
(ADCCC; {\tts hea-www.harvard.edu/adccc}).
The Canadian Astronomy Data Center (CADC; {\tts cadcwww.dao.nrc.ca})
includes archives of the CFHT, JCMT and UKIRT telescopes on Hawaii.
The ESO and ST-ECF Science Archive Facility at {\tts archive.eso.org}
offers access to data obtained with the ``New Technology Telescope'' (NTT),
and to the catalogue of spectroscopic plates obtained at ESO telescopes
before 1984.
The ``Hubble Data Archive'' at {\tts archive.stsci.edu} includes
the HST archive, the International Ultraviolet Explorer (IUE) archive
and the VLA FIRST survey.
The La Palma database at {\tts archive.ast.cam.ac.uk/ingarch} contains
most observations obtained with the Isaac Newton group of telescopes
of RGO.
At the National Optical Astronomy Observatory
(NOAO; {\tts www.noao.edu/archives.html}) most data from CTIO and KPNO
telescopes are now being saved, and are available by special permission
from the Director.
Data from several high-energy satellite missions may be retrieved from
HEASARC ({\tts heasarc.gsfc.nasa.gov}), and from LEDAS (\S\ref{pseudo}).
For results of the Space Astro\-metry Mission {\it HIPPARCOS}
see {\tts astro.estec.esa.nl/Hipparcos/hipparcos.html} \linebreak[4]  or
{\tts cdsweb.u-strasbg.fr}.

Images from many surveys of the whole or most of the sky can
be retrieved from the SkyView facility ({\tts skyview.gsfc.nasa.gov})
at Goddard Space Flight Center (GSFC).
Documentation of these surveys is available at
{\tts skyview.gsfc.nasa.gov/cgi-bin/survey.pl}. Overlays of these
surveys with either contours from another survey or objects from a
large set of object catalogues may be
requested interactively.

Until late 1997 there was no tool which unified the access to
the multitude of existing observatory archives. A serious approach
to this goal has been made within the ``AstroBrowse'' project
({\tts \verb*Csol.stsci.edu/~hanisch/astrobrowse_links.htmlC}).
This resulted in the ``Starcast'' facility
{\tts faxafloi.stsci.edu:4547/starcast} allowing one to find (and retrieve)
relevant data (photometric, imaging, spectral or time series) on a given object
or for a given region of sky, in any range of the electromagnetic spectrum
from ground- or space-based observatories.  The HEASARC ``AstroBrowser''
({\tts legacy.gsfc.nasa.gov/ab}) provides an even wider scope, including
astronomical catalogues and {\tt VizieR} at the same time.
The `` Multimission Archive at STScI'' (MAST;
{\tts archive.stsci.edu/mast.html}) currently combines the archive of HST,
IUE, Copernicus, EUVE, UIT, plus that of the FIRST radio survey
and the DSS images. Its interface with astronomical catalogues
allows one to query e.g.\ which high-redshift QSOs have been observed
with HST, or which Seyfert galaxies with IUE.

An ``Astronomy Digital Image Library'' (ADIL) is available from
the National Center for Supercomputing Applications (NCSA) at
{\tts imagelib.ncsa.uiuc.edu/imagelib}.

\section{Digital Optical Sky Surveys, Finding Charts, \& Plate Catalogues} \label{dssetc}

The first Palomar Observatory Sky Survey (POSS-I) was taken on glass plates
in red (E) and blue (O) colour from 1950 to 1958. While glass copies are
less commonly available, the printed
version of POSS-I provided the first reference atlas of the whole sky
north of $-$30\deg\ declination down to $\sim$20\,mag. Together
with its southern extensions, provided 20--30 years later by ESO in B,
and by the UK Schmidt Telescope (UKST) in B$_J$, it
was the basic tool for optical identification of non-optical objects.
However, reliable optical identification required a positional accuracy
on the order of a few arcsec. A common, but not too reliable, tool for this
were transparent overlays with star positions taken from the SAO star
catalogue. Major limitations are the low density of stars at
high Galactic latitudes and differences in the scale and
projection between the transparency and the Palomar print or plate.
For higher accuracy than a few arcsec
the use of a plate measuring machine was required for triangulation of
fainter stars closer to the object in question, but such machines were
only available at a few observatories.

\begin{figure*}[!ht] 
\vspace*{2mm}

\hspace*{-1mm}
\mbox{
\epsfig{file=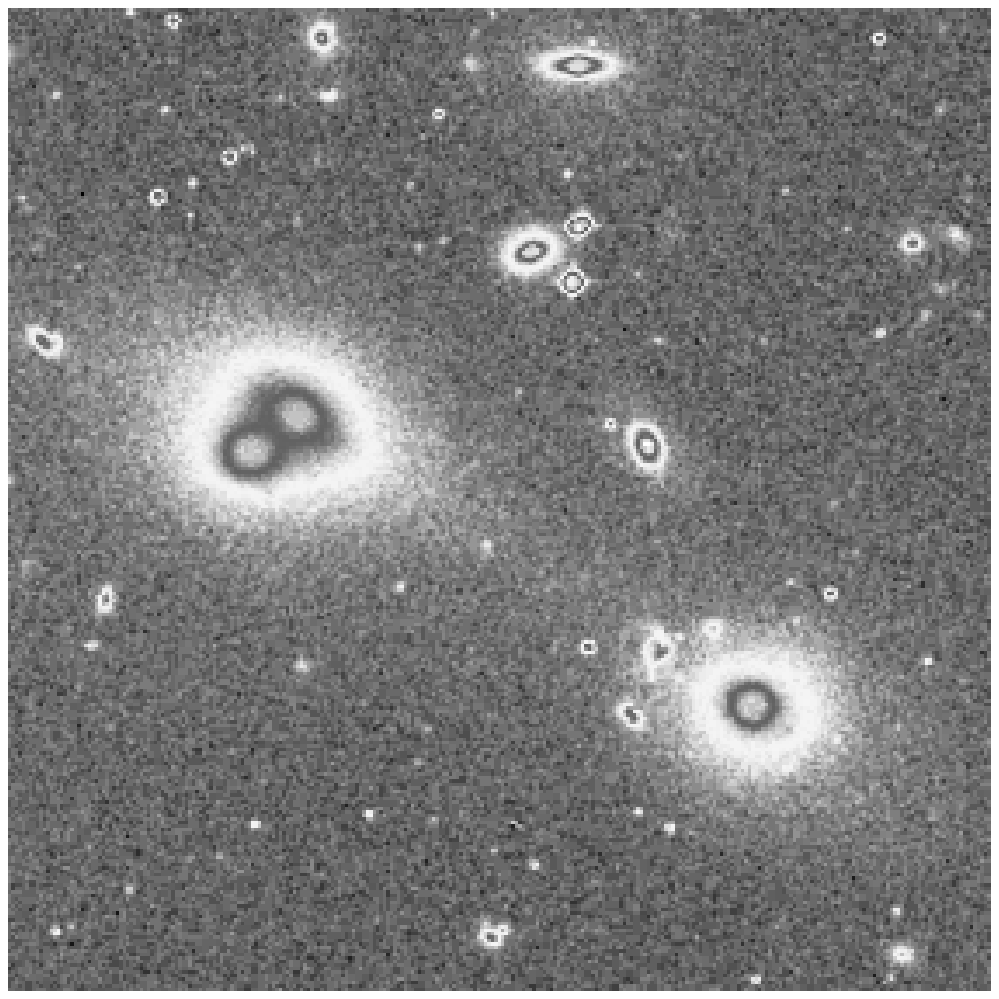,width=6.5cm}
\hspace*{1mm}
\epsfig{file=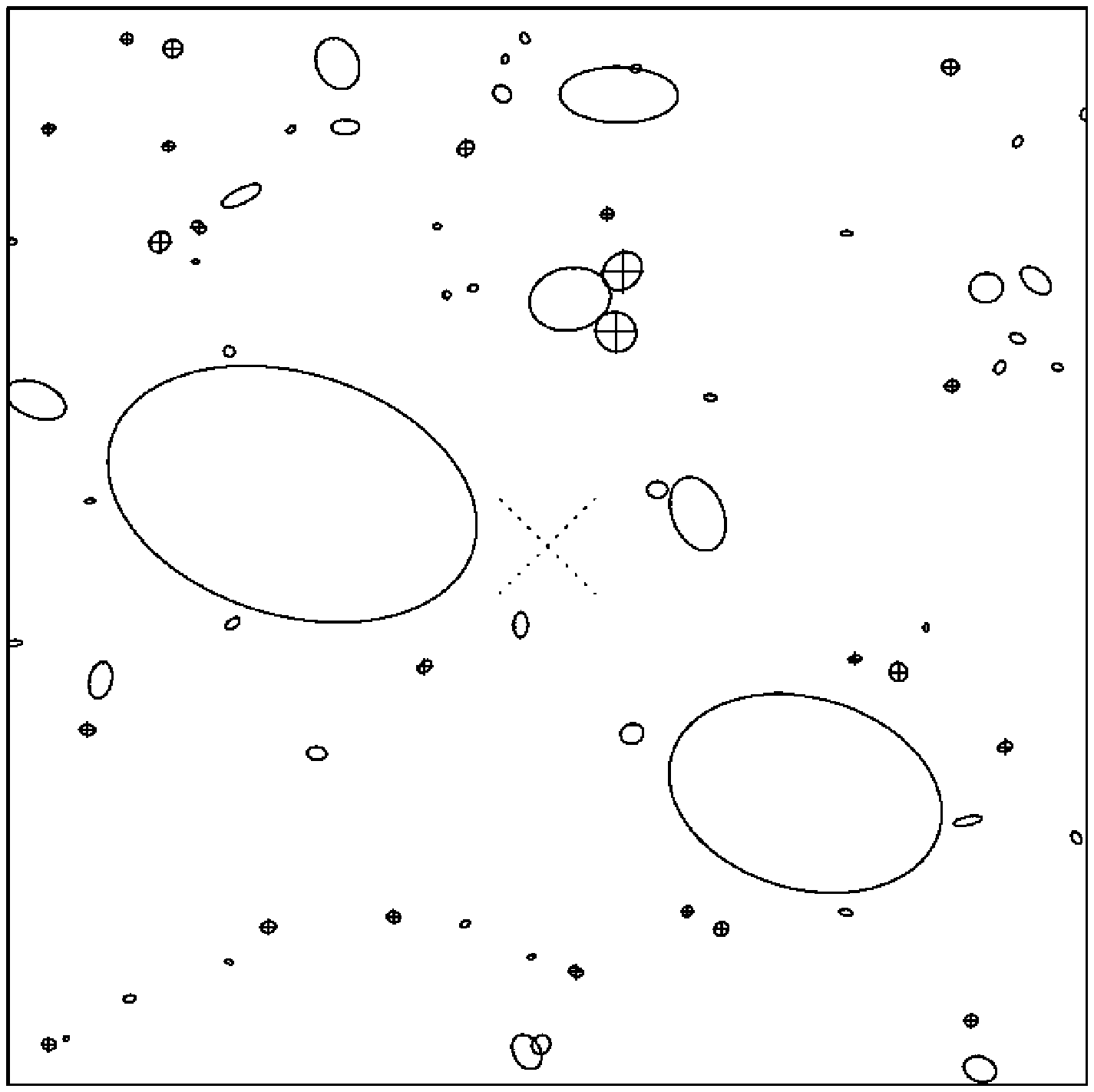,width=6.6cm}
\vspace*{0.5cm}}

\vspace*{0.5cm}
\hspace*{-1.5mm}
\mbox{
\epsfig{file=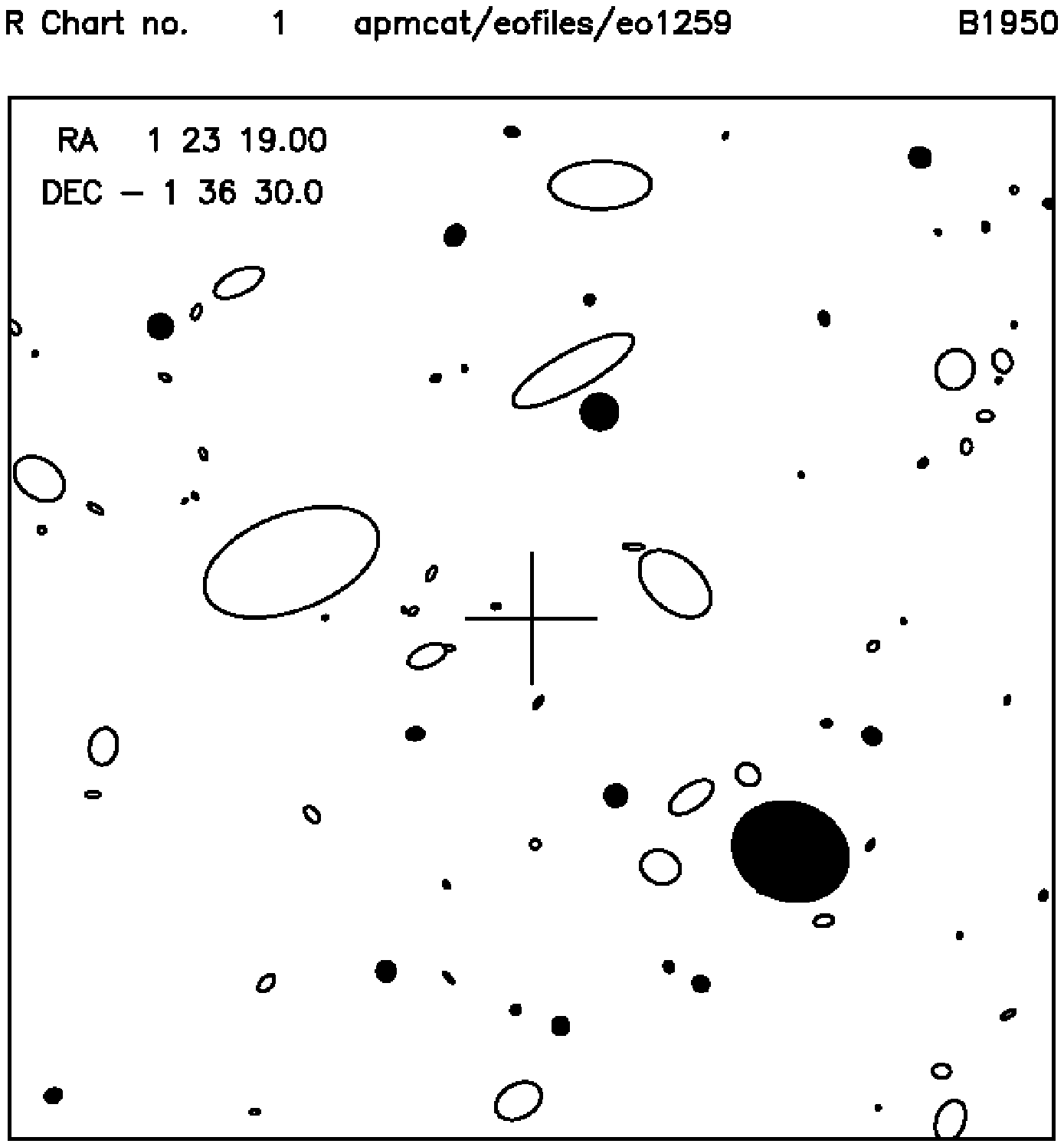,width=6.57cm}
\hspace*{0.5mm}
\epsfig{file=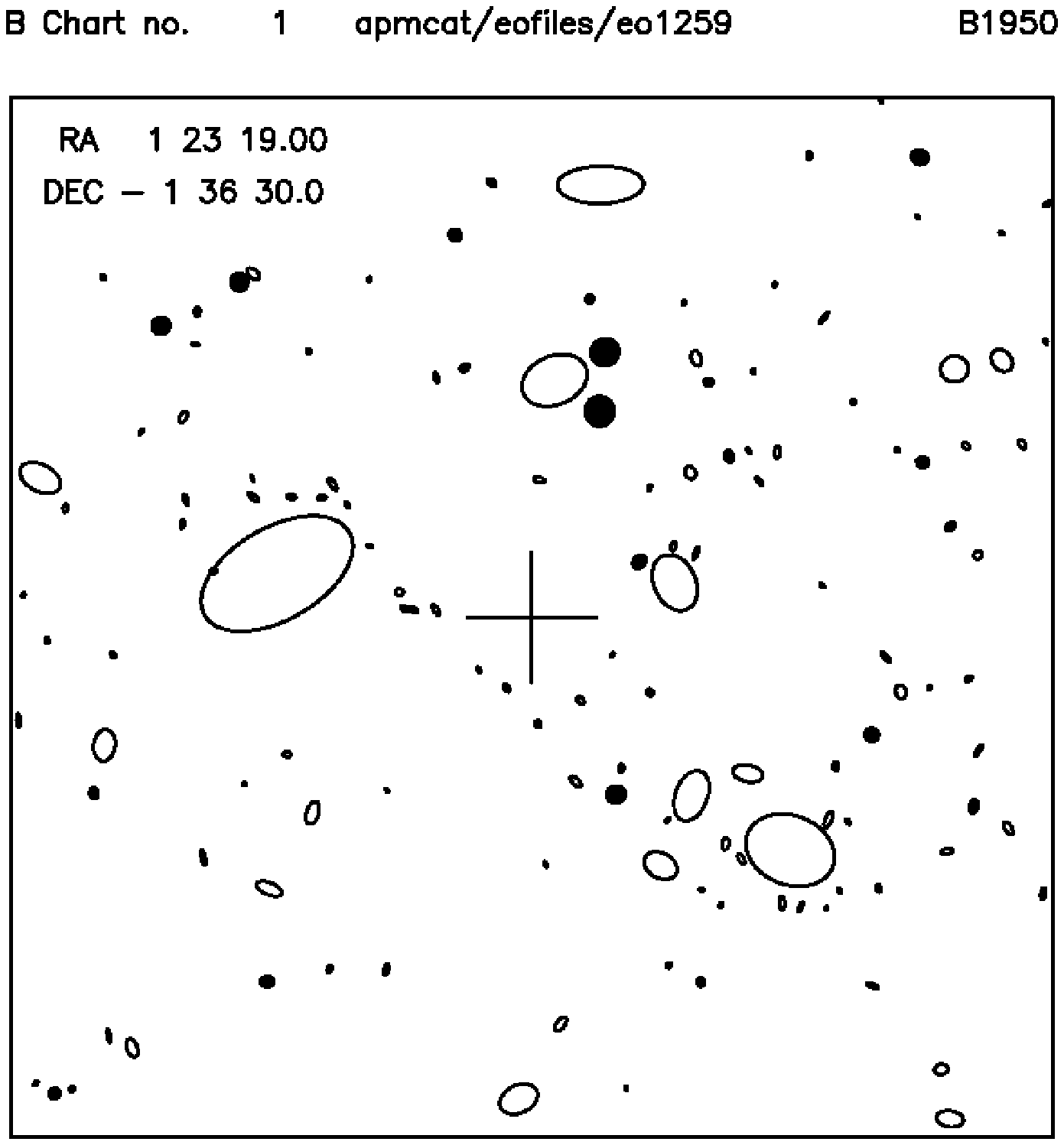,width=6.57cm}
}

\caption{Digitized Sky Survey (DSS) images versus finding charts
from object catalogues, for an 8$'\times$8$'$ region in the
core of the galaxy cluster Abell~194.
Upper left: DSS image of POSS-I red plate from {\tt SkyView};
upper right: COSMOS finding chart from B$_J$ plate (open ellipses
are galaxies, star-like objects are crossed);
lower panels: APM charts from POSS-I, red (R)
and blue (B) plate (filled symbols are star-like, open ones
are galaxy-like, and the central cross is 1~arcmin wide).
Note the different classification of objects in APM and COSMOS and
in the R and B scans of APM. Several multiple objects, clearly
separated on the DSS, are blended into single, often elongated objects
in the APM or COSMOS catalogues.}  \label{dssfig}
\end{figure*}

Curiously, it was mainly the necessity, around 1983, to prepare the
HST Guide Star Catalog (GSC;~{\tts www-gsss.stsci.edu/gsc/gsc.html}),
which led to a whole new
Palomar Sky survey (the ``quick-V'') with shorter exposures
in the V band, which was then fully digitized at STScI with 1.7$''$ pixel size
to extract guide stars for the HST. Later the deeper red plates of POSS-I were
also scanned at the same resolution to provide an image database
of the whole northern sky.  For the first time almost the whole sky was
available with absolute accuracy of $\sim$1$''$, but owing to the
sheer volume the pixel data were accessible only to
local users at STScI during the first years.  By the time that they had
been prepared for release on 102 CD-ROMs and sold by the Astronomical Society
of the Pacific (ASP), the Internet and the WWW had advanced to a point where
small extractions of these pixel data could be accessed remotely.
Other observatories also employed plate-scanning machines
to scan POSS, ESO and UKST surveys at even finer pixel sizes, and some
catalogues were prepared that contained several 10$^8$ objects detected
on these plates. Such catalogues usually include a classification of the
object into stars, galaxies or ``junk'' (objects which fit into neither
class and may be artefacts). However, such classifications have a
limited reliability. They should not be taken for granted, and it is
wise to check the object by visual inspection on the plates (or prints
or films), or at least on the digitized image. It is important to
distinguish between these different media:
the glass plates may show objects of up to $\sim$1\,mag fainter
than are visible on the paper prints.
Thus the pixel data, being digitized from the glass plates,
may show fainter objects than those visible on the prints.
Moreover, they offer absolute positional accuracy of better
than 1$''$. On the other hand, the pixel size of the standard DSS (1.7$''$)
represents an overriding limitation in deciding on the morphology
of faint (i.e. small) objects.
Eventually we have the ``finding charts'' which are merely sketches of
all the objects extracted from the image, plotted to scale, but with
artificial symbols representing the object's magnitude, shape,
orientation, etc. (usually crosses or full ellipses for stars, and open
ellipses for non-stellar objects).  They should not be taken
as a true image of the sky, but rather as an indication of the presence
of an optical object at a given position, or as an accurate orientation
indicator for observers (see Fig.~\ref{dssfig} for an example).

In what follows I present a quick guide to the various data products
which can be freely accessed now.
The Royal Observatory Edinburgh (ROE) offers comprehensive information on the
status of ongoing optical sky surveys at {\tts www.roe.ac.uk/ukstu/ukst.html}
(go to the ``Survey Progress'' link).  Other places to watch for such information
are ``Spectrum'' (the RGO/ROE Newsletter),
the ESO Messenger ({\tts www.eso.org/gen-fac/pubs/messenger}),
the STScI Newsletter ({\tts www.stsci.edu/stsci/newsletters/newsletters.html}),
the Anglo-\linebreak[4] Australian Observatory (AAO) Newsletter
({\tts www.aao.gov.au/news.html}) and the Newsletter of the
``Working Group on Sky Surveys'' (formerly ``WG on Wide Field Imaging'')
of IAU Commission 9
(chaired by Noah Brosch, email {\tts noah@stsci.edu}, a URL at
{\tts http://www-gsss.stsci.edu/iauwg/welcome.html} is in preparation).
See also the chapter by D.~Golombek for a summary of plate digitizations
available at the STScI.

\subsection{Digitized Sky Survey Images} \label{dssimag}

Digitizations of the POSS E plates in the northern sky and UKST B$_J$
plates in the southern sky
are available from {\tt SkyView} at URL {\tts skyview.gsfc.nasa.gov}.
For the vast majority of astronomical institutions without a CD-ROM
juke box, this web service allows much easier access to
the DSS than working locally with the set of 102 CD-ROMs of the ASP.
Batch requests for large lists of DSS extractions can be formulated
from the command line using server URLs based on {\tt perl} scripts (see
{\tts skyview.gsfc.nasa.gov/batchpage.html} or
{\tts \verb*Cwww.ast.cam.ac.uk/~rgm/first/collab/first_batch.htmlC}).
Several other sites offer the standard DSS through their servers, e.g.\
the CADC at
{\tts cadcwww.dao.nrc.ca/dss/dss.html}, ESO at
{\tts archive.eso.org/dss/dss}, or {\tt Skyeye} at Bologna
({\tts www.ira.bo.cnr.it/skyeye/gb}), the University of Leicester in UK
({\tts ledas-www.star.le.ac.uk/DSSimage}), and NAO in Japan at
{\tts dss.mtk.nao.ac.jp}. Each of these offer a
slightly different ``look and feel'',
e.g. CADC offers absolute coordinates labels around the charts (but
only in J2000), SkyEye offers simultaneous extraction of objects from DIRA2
catalogues (\S\ref{pseudo}) and an easy batch request for charts via email, etc.
A comparison of performance and speed between these DSS servers
can be found at
{\tts \verb*Cstar-www.rl.ac.uk/~acc/archives/archives.htmlC}
(see \linebreak[4] \cite{acc98}).

Meanwhile almost all of the ``second-epoch survey'' (SES)
plates have been taken: the northern POSS-II
({\tts www.eso.org/research/data-man/poss2}) at Palomar,
and the southern UKST SES-R survey at AAO
({\tts www.roe.ac.uk/ukstu/ukst.html}).
Most of these have been digitized. Early in 1998
the STScI server ({\tts archive.stsci.edu/dss/dss\_form.html})
was the only one offering the second-epoch surveys (POSS-II or UKST R),
if available, and otherwise POSS\,I or UKST B$_J$ surveys.

A digitization of the POSS-I E- and O-survey plates was also performed with the
Automated Plate Scanner (APS) at the Astronomy Department of the University of
Minnesota. Only those plates (E and O) with $|b|>$20\deg\ have been scanned.
The APS uses a flying laser spot to record the transmission of each
plate only above $\sim$65\% of the background transmittance as determined
in an initial low-resolution scan. This compromise was needed to achieve
the small pixel size of 0.3$''$ for the final images which
contain only signal above background, i.e. their background is black.
They can be retrieved in FITS format from {\tts aps.umn.edu/homepage.aps.html}.

\subsection{Object Catalogues and Finding Charts from DSS} \label{ocatdss}

The HST Guide Star Catalog (GSC) was the first all-sky catalogue of optical
objects extracted from plate digitizations. For declinations north of $+$3\deg
the Palomar ``Quick-V'' (epoch 1982) plates of 20\,min exposure were used.
For the south the 50--75\,min exposures of the SERC B$_J$ survey
(epoch $\sim$1975) and its equatorial extension (epoch $\sim$1982) were
used (see {\tts www-gsss.stsci.edu/gsc/gsc12/description.html}).
The GSC contains $\sim$19\,10$^6$ objects in the range 6--15~mag. Most of them are
stars, but an estimated 5\,10$^6$ galaxies are present as well.
The positional accuracy  has been improved to better than 0.4$''$ in version
1.2.  Note, however, that this catalogue is not magnitude-limited, but that
the selection of stars has been carried out so as to provide a homogeneous
{\it density} of guide stars over the sky.

The {\it Automated Plate Measuring Machine} (APM) is located at the
Institute of Astronomy, Cambridge, UK and has been used to prepare object
catalogues from Sky Survey plates at high Galactic latitudes
($|b|>$20\deg), see e.g. \cite{apm96}.
Both colours of the POSS-I survey plates were scanned and the
objects cross-identified, so that colour information is available
for a matched object catalogue of well over
100 million objects down to m=21.5 in blue (O) and m=20 in red (E).
For the southern sky the glass plates of the UKST B$_J$ and later the
UKST SES-R survey have been scanned, with limiting magnitudes of 22.5 in
B$_J$ and 21 in R.
All plates were scanned at 0.5$''$ scan interval and a scanning resolution of
1$''$. The pixel data of the scans are not available, and
no copies of the entire catalogue are distributed.
Both the northern hemisphere catalogue ($\delta > -$3\deg\ ),
and the southern hemisphere catalogue based on UKST B$_{J}$ and SES-R plates
($\sim$50\% complete) are available for routine interrogation at
{\tts \verb*Cwww.ast.cam.ac.uk/~apmcatC}. The URL
{\tts www.aao.gov.au/local/www/apmcatbin/forms} offers a standalone
client program in C ({\tt apmcat.c}) which allows queries for large sets
of finding charts and object lists from the command line.
The catalogues can also be accessed from a captive account
({\tts telnet 131.111.68.56}, login as {\tts catalogues} and follow
the instructions).

COSMOS (COordinates, Sizes, Magnitudes, Orientations, and Shapes) is a
plate scanning machine at the Royal Observatory Edinburgh, which was
used to scan the whole southern sky ($\delta<+2.5$\deg) from
the IIIa-J and Short Red Surveys, and led to an object
catalogue of several hundred million objects
(\cite{drinkwater95}).  Public access to the
catalogue is provided through the Anglo-Australian Observatory (AAO)
at {\tts www.aao.gov.au/local/www/surveys/cosmos},
and through the Naval Research Laboratory (NRL) at
{\tts xweb.nrl.navy.mil/www\_rsearch/RS\_form.html}.
The AAO facility requires the user to register and obtain a password.
During a {\tt telnet} session the user may either input coordinates
on the fly, or have them read from a file previously transferred
(via {\tt ftp}) to the public AAO account, and create charts and/or object lists
on various different output media. The user has to transfer the output
back to the local account via {\tt ftp}. This disadvantage is balanced by the
possibility of extracting large amounts of charts for big cross-identification
projects. Charts may be requested in stamp-size format resulting in
PostScript files containing many charts per page.

The US Naval Observatory (USNO) has scanned the POSS I E- and O-plates (for plate
centres with $\delta\ge -30$\deg) and the ESO-R and SERC-J plates
(centred at $\delta\le -35$\deg) with the ``Precision Measuring Machine'' (PMM).
A scan separation of 0.9$''$ was used (i.e. finer than that of the STScI
scans for DSS), and object fitting on these
images resulted in the USNO-A1.0 catalogue of 488,006,860 objects
down to the very plate limit (limiting mag O=21, E=20, J=22, F=21).
Objects were accepted only if present to within 2$''$ on both
E- and O-plates, which implies an efficient rejection of plate faults,
but also risks losing real, faint objects with extreme colours.
This catalogue is available both as a set of 10 CD-ROMs and
interactively at {\tts psyche.usno.navy.mil/pmm}.
Client programs at CDS (\S\ref{catal}) and
ESO ({\tts archive.eso.org/skycat/servers/usnoa})
allow extraction of object lists of small parts of the sky
very rapidly from the command line.
There are plans to produce a USNO-B catalogue, which will combine POSS-I and
POSS-II in the north, UKST B$_J$, ESO-R, and AAO-R in the south, and will
attempt to add proper motions and star/galaxy separation fields to
the catalogue.

The images from the APS scans of POSS-I (see above) have been used to
prepare a catalogue of $\sim10^9$ stellar objects and $10^6$ galaxies
detected on {\it both} E and O plates.
Extractions from this catalogue may be drawn from~
{\tts aps.umn.edu/homepage.aps.html},
or from the ADS catalogue service
at {\tts adscat.harvard.edu/catalog\_service.html}.

\subsection{Catalogues of Direct Plates}

While the object catalogues mentioned above were drawn from
homogeneous sky surveys, there are almost 2 million wide-field
photographic plates archived at individual observatories around
the world (see the Newsletters of the IAU Commission 9
``Working Group on Sky Surveys'').
To allow the location and retrieval of such plates for possible inspection
by eye or with scanning machines, the ``Wide-Field Plate Database'' (WFPDB)
({\tts www.wfpa.acad.bg}) is being compiled and maintained at
the Institute of \linebreak[4] Astronomy of the Bulgarian Academy of Sciences.
As described by \cite{wfpa97}, the WFPDB offers metadata for
currently $\sim$330,000 plates from 57 catalogues
(see {\tts \verb*Cwww.wfpa.acad.bg/~listC}) collected from more than
30 observatories.  Since August 1997 the WFPDB may be searched
as catalogue {\tt VI/90} via the CDS {\tt VizieR} catalogue browser
({\tts vizier.u-strasb.fr/cgi-bin/VizieR}).

\subsection{An Orientation Tool for the Galactic Plane}

The ``Milky Way Concordance'' is a graphical tool to create
charts with objects from catalogues covering the
Galactic Plane, as described by \cite{barnes97}, available at the URL
{\tts \verb*Ccfa-www.harvard.edu/~peterb/concordC}. Currently 17 catalogues
including H\,II regions, planetary and reflection nebulae, dark clouds,
and supernova remnants are available to create colour-coded finding charts
of user-specified regions.

\subsection{Future Surveys}

The All Sky Automated Survey
(ASAS; {\tts \verb*Csirius.astrouw.edu.pl/~gp/asas/asas.htmlC}) is a project
of the Warsaw University Astronomical Observatory (\cite{pojman97}).
Its final goal is the photometric monitoring of $\sim$10$^7$ stars
brighter than 14\,mag all over the sky from various sites distributed
over the world. The first results on variable stars found in $\sim$100
square degrees have become available at the above URL.

While current Digitized Sky Surveys are all based on photographic material
digitized off-line after observing, the future generation of optical Sky Surveys
will be digital from the outset, like the ``Sloan Digital Sky Survey'' (SDSS;
{\tts www-sdss.fnal.gov:8000}, \linebreak[4] \cite{margon98}). 
The SDSS will generate deep (r$'$=23.5\,mag)
images in five colours (u$'$, g$'$, r$'$, i$'$, and z$'$) of $\pi$ steradians
in the Northern Galactic Cap ($|$b$|>+$30\deg).
The SDSS will be performed in drift-scan mode over a period of five years.
A dedicated 2.5\,m telescope at Apache Point Observatory
(NM, USA; {\tts www.apo.nmsu.edu}), equipped with a mosaic of 5$\times$6
imaging CCD detectors of 2048$^2$ pixels will allow a
uniquely large 3\deg\  field of view.
Selected from the imaging survey, 10$^6$ galaxies (complete to r$'\sim$18\,mag)
and 10$^5$ quasars (to r$'\sim$19\,mag) will be observed spectroscopically.
The entire dataset produced during the course of the survey will be
tens of terabytes in size.  The SDSS Science Archive
({\tts tarkus.pha.jhu.edu/scienceArchive}) will eventually contain several
10$^8$ objects in five colours, with measured attributes, and associated
spectral and calibration data.
Observations are due to begin in 1998, and the data will be made available
to the public after the completion of the survey.

\section{Bibliographical Services} \label{bibserv}

Long before the Internet age, abstracts of the astronomical literature
were published annually in the {\it Astronomischer Jahresbericht}
by the Astronomisches Rechen-Institut in Heidelberg
({\tts www.ari.uni-heidelberg.de/publikationen/ajb}).
The series started in 1899, one year after the first issue of {\it Science
Abstracts} was published, the precursor of INSPEC (\S\ref{absserv};
cf. {\tts www.iee.org.uk/publish/inspec/inspec.html}).
Abstracts of many papers which originally
did not have an English abstract were given in German. Since 1969 its
successor, the {\it Astronomy \& Astrophysics Abstracts} (AAA), published
twice a year, have been THE reference work for astronomical bibliography.
The slight drawback that it appears about 8 months after the end of
its period of literature coverage is compensated by its impressive
completeness of ``grey literature'', including conference proceedings,
newsletters and observatory publications.
Until about 1993, browsing these books was about the only means for
bibliographic searches ``without charge'' (except for the cost of the
books). In 1993 NASA's ``Astrophysics Data System'' (ADS)
Abstract Service with initially
160,000 abstracts became accessible via telnet. After a few months
of negotiation about public accessibility outside the US,
the service was eventually put on the WWW in early 1994,
with abstracts freely accessible to remote users world-wide.
Shortly thereafter they turned into (and continue to be) the most popular
bibliographic service in astronomy (see \cite{kurtz96,eichhorn98}).

Upon the announcement during IAU General Assembly XIII (Kyoto, Japan,
Aug.~1997) that AAA is likely to stop publication at the end of 1998,
some Astronomy librarians compared the completeness of AAA with that
of ADS and INSPEC (\S\ref{absserv}). The results show that,
in particular, information about conference proceedings and observatory reports
is missing from ADS and INSPEC. After the demise of AAA, ADS would be the
{\it de facto} bibliography of astronomy literature, and there is a danger
that it will not be as complete as AAA (see
{\tts www.eso.org/libraries/iau97/libreport.html} for a discussion).
It would indeed be to the benefit of all astronomers if some day all abstracts
from {\it Astronomischer Jahresbericht} and AAA (covering 100 years!)
became available on the Internet (see \S\ref{absserv} for ARIBIB).

Many bibliographic services in astronomy use a 19-digit reference code or
``refcode'' (\cite{schmitz95}, see e.g.\
{\tts cdsweb.u-strasbg.fr/simbad/refcode.html}).
They have the advantage of being unique, understandable to the human eye,
and may be used directly to resolve the full reference and to see their
abstracts on the web. Lists of refcodes are also maintained by
NED and ADS ({\tts adsabs.harvard.edu/abs\_doc/journal\_abbr.html}). Note,
however, that ADS calls them ``bibcodes'', and that for less common
bibliographic sources occasionally these may differ from CDS refcodes.
Unique bibcodes do not exist as yet for proceedings
volumes and monographs, but work is under way in this area.

\subsection{Abstract and Article Servers} \label{absserv}

NASA's ADS Abstract Service ({\tts adsabs.harvard.edu/abstract\_service.html})
is offered at the Center for Astrophysics (CfA)
of the Smithsonian Astrophysical Observatory (SAO). It
goes back to several 10$^5$ abstracts prepared by NASA's`` Scientific and
Technical Information'' Group (STI) since 1975. Note that the latter
abstracts may not be identical with the published ones and that complete
coverage of the journals is not guaranteed. Since 1995 most of the
abstracts are being received directly from the journal editors,
and coverage is therefore much more complete.
The service now contains abstracts from four major areas which need
to be searched separately: Astronomy ($\sim$380,000 abstracts), Instrumentation,
Physics \& Geophysics and LANL/SISSA {\tts astro-ph} preprints (\S\ref{ppserv}).
The preprints expire 6 months after their entry date.
The four databases combined offer over 1.1 million references.
The service is also useful to browse contents of recent journals using
the {\tts BIBCODE QUERY} or {\tts TOC QUERY} (\S\ref{tocs}) links.
Its popularity is enormous: it was accessed
by $\sim$10,000 users per month, and about 5 million references
per month were returned in response to these queries in late 1997.
It has mirror sites in Japan ({\tts ads.nao.ac.jp}) and France
({\tts cdsads.u-strasbg.fr/ads\_abstracts.html}).

The ADS provides very sophisticated search facilities, allowing one to filter by
author, by title word(s) or words in the abstract, and even by object name,
albeit with the silent help of NED and SIMBAD.
The searches can be tuned with various weighting schemes and the resulting
list of abstracts will be sorted in decreasing order of relevance
(see {\tts adsdoc.harvard.edu/abs\_doc/abs\_help.html} for
an extensive manual).
Each reference comes with links (if available) to items like
(C) citations available (references that cite that article),
(D) data tables stored at CDS or ADC, (E) electronic versions of the
full paper (for users at subscribing institutions), (G) scanned version
of the full paper, (R) references cited by that article, etc.
Links between papers (citations and references) are gradually being
completed for older papers. Citations are included for papers published
since 1981 and were purchased from the ``Institute for Scientific
Information'' (ISI), see below.
When the recognition of the full text from the scanned images has
been completed in a few years (see below), ADS plans to build
its own R and C links.


The ADS also employed page scanners to convert printed pages
of back issues of major astronomical
journals into images (``bitmaps'') accessible from the web.
Early in 1995 the ADS started offering this ``Article Service''
at {\tts adsabs.harvard.edu/article\_service.html}.
Images of over 60,000 scanned articles are now on line, and over 250,000
pages were retrieved monthly in 1997. The intention is to prepare
page scans back to volume~1 for all major journals. Note, however, that
these are images of printed pages and not ASCII versions of the full text.
Eventually the full article database will be about 500 Gbyte of data.
The images of  printed pages may eventually be converted into ASCII text
via OCR, but currently this exercise is estimated to take a full year of
CPU time (excluding the subsequent effort in proof-reading and correcting
the OCR result). No full-text search features are available as yet
from ADS even for recent articles.

ARIBIB is a bibliographic database with information given in the
printed bibliography ``Astronomy and Astrophysics Abstracts''
(AAA) by the Astrononomisches Recheninstitut (ARI) in Heidelberg, Germany.
Currently, at~ {\tts www.ari.uni-heidelberg.de/aribib}
references to the literature of 1983--1997 (Vols.\ 33--68 of AAA)
are freely available, while abstracts may be retrieved only by 
subscribers of the printed AAA.
The ARI intends to prepare abstracts of older literature in a 
machine-readable format, by scanning earlier volumes of AAA and 
``Astronomischer Jahresbericht''. 

The UnCover database of authors and titles of scientific papers
(see \S\ref{tocs} for details) may also be used for keyword searches,
although it does not offer abstracts.

There are numerous well-established commercial bibliographic database
services which charge for access.  The use of these systems in astronomy
has been reviewed by Davenhall (1993) and \cite{michold95}, and
\cite{thomas97} gives a more general overview.  Typically these databases cover a
range of scientific and engineering subjects and none of
them is specifically astronomical.  This has the disadvantage that
more obscure astronomical journals are not covered, but the advantage
that astronomical papers in non-astronomical publications will be included.
The Institution of Electrical Engineers (IEE) in the UK produces
INSPEC ({\tts www.iee.org.uk/publish/inspec/inspec.html}) which is
the main English-language commercial bibliographic database
covering physics (including astrophysics), electrical engineering,
electronics, computing, control and information technology.
It currently lists some 5 million papers and reports, with over
300,000 new entries being added per year, and it covers the
main astronomical journals.
An abstract is usually included for each entry.
Another important bibliographic database is the Science Citation
Index (SCI) produced by the Institute for Scientific Information Inc.~(ISI;
{\tts www.isinet.com}) in USA.  The SCI contains details drawn from
over 7500 journals and ({\it via} the {\it Index to Scientific and
Technical Proceedings}, ISTP) over 4200 conferences per year.
While the SCI does not contain abstracts, it offers
cross-references to all the citations in each paper that is included,
a unique and extremely valuable feature.  Often commercial
bibliographic services are not accessed directly, but rather through a
third-party vendor.  Typically such a vendor will make a number of
bibliographic databases available, having homogenized their appearance
and customized their contents to a greater or lesser extent.  There is
a number of such vendors; one example is the ``Scientific and Technical
Information Network'' (STN; {\tts www.cas.org/stn.html}) which includes
links to about 200 databases.

\subsection{Preprint Servers}  \label{ppserv}

\begin{figure*}[!ht]
\hspace*{6mm}
\epsfig{file=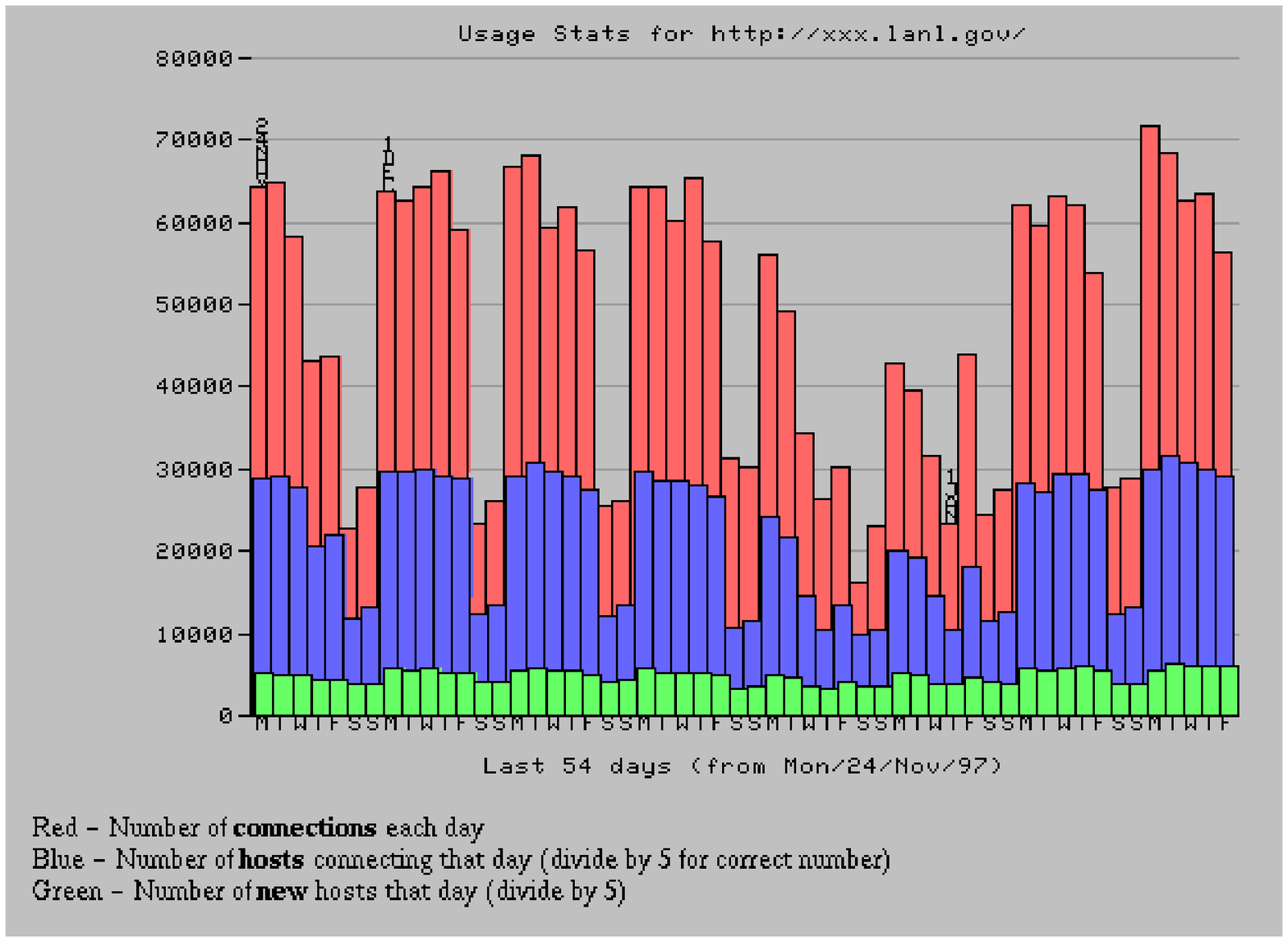,width=12.cm}
\caption{Daily accesses to the LANL preprint server. 
The regular 2-day dips are weekends, while the wider dips are 
due to Thanksgiving, Christmas and New Year 1997/98.} \label{lanluse}
\end{figure*}

The availability of electronic means has reduced the delay between acceptance and
publication of papers in refereed journals from 6--10 down to 3--6 months
(or even 1 month in case of letters or short communications),
largely because authors prepare their own manuscripts electronically
in the formatting requirements of the journals. Nevertheless, for conference
proceedings the figure remains between 10 and 20 months. Such a delay
did (and still does) crucially affect certain types of research. Thus,
for several decades, the ``remedy'' to this delay has been
a frequent exchange of preprints among astronomers,
and until very recently, the preprint shelves used to be the most
frequented areas in libraries.
This situation has gradually changed since April 1992, when both SISSA
(International School for Advanced Studies, Trieste, Italy) and LANL
(Los Alamos National Laboratory, USA)
started to keep mirror archives of electronically submitted preprints
({\tts xxx.lanl.gov}). In the first years of their existence, preprints
were dominant in the fields of theoretical cosmology and particle
physics, and about 35 {\tts astro-ph} preprints were submitted monthly
in mid-1993.  The popularity of this service has increased impressively
since then: over 60,000 {\it daily} accesses to {\tts xxx.lanl.gov} from
6,000 different hosts (see Fig.\,\ref{lanluse}), and
about 300 preprints submitted per month in 1997/8 only
for {\tts astro-ph} (and $\sim$1800 altogether), with a fair balance
between all parts of observational and theoretical astrophysics;
mirror sites have been installed in 12 other countries (Australia,
Brazil, China, France, Germany, Israel, Japan, Russia, South Korea,
Spain, Taiwan, and the UK, see {\tts xxx.lanl.gov/servers.html}).
A further mirror site is under construction in India. References to
electronic preprints from LANL/SISSA are made more and more frequently in
refereed journals, and the LANL/SISSA server also provides links to
citations to, and references from, their electronic preprints to other
preprints of the same collection.

You may subscribe to receive a {\it daily} list of new preprints.
To avoid this flood of emails you could rather send an email
to {\tts astro-ph@xxx.lanl.gov} just after the end of month mm of
year yy, with subject ``list {YY}{MM}'' , e.g.\ ``list 9710'',
to receive a full list of preprints archived in October 1997.
To get the description of all the functionalities of the preprint
server (e.g. how to submit or update your own preprints), send an
email to the above address with ``{\tts Subj:}\,{\tts get bighelp.txt}''.
If you want to check for the presence of a certain preprint, send an
email with ``{\tts Subj:}\,{\tts find keyword}''~, where {\tts keyword} is an author
or part of the title of the paper. You need to add a year if you want to
search further back than the default of one year from the current date.
Once you have located the preprint(s) that you are interested in, the most
efficient (least time-consuming) method to obtain a copy is to request
it by its sequence number {\small {YY}{MM}{NNN}}, via email to the same address with
{\tts Subj:}\,{\tts get {YY}{MM}{NNN}}. It comes in a self-extracting uuencoded file
which you need to save in a file, say {\small XYZ}, strip off the mail header,
and just execute it with the command {\tts csh XYZ}. A common shortcoming is that
authors sometimes do not include all necessary style files with their papers.
Generally these can be obtained separately from the server by email
with {\tts Subj:}~{\tts get whatever.sty}. Another solution to this problem is
to use the web interface at {\tts xxx.lanl.gov/ps/astro-ph/{YY}{MM}{NNN}},
where (unlike the email or {\tt ftp} service)
the complete PS files of papers are indeed available (and are in fact
created on the fly upon request).

Submission of a preprint to the LANL/SISSA server is today the most
efficient way of world-wide ``distribution'' without expenses for
paper or postage, since the preprint will be available to the entire
community within 24 hours from receipt, provided it passes some technical
checks of file integrity and processability.  Note, however, that
electronic preprint servers contain papers in different stages
of publication: accepted, submitted to a refereed journal or to appear
in conference proceedings. Occasionally the authors do not explain the
status of the paper, and the preprint server may have been
the only site to which the paper has ever been submitted.
Of course, a reference made to a preprint of work that was never
published should be regarded with caution in case it was subsequently
rejected by a journal.  Note also that
only a fraction of all papers published in refereed journals (perhaps
30--50\% now) is available from LANL/SISSA prior to publication.
Many other preprints are offered from web pages of individual researchers
or from institutional pages.
A substantial collection of links to other sources of preprints is
provided at {\tts www.ucm.es/info/Astrof/biblio.html\#preprints}.

The International Centre for Theoretical Physics in Trieste, Italy,
provides the ``One-Shot World-Wide Preprints Search'' at
{\tts www.ictp.trieste.it/indexes/preprints.html}.

You can also check a list of ``Papers Submitted to
Selected Astrononomical Publications'' at
{\tts www.noao.edu/apj/ypages/yp.html}. A list of papers
submitted to the {\it Astronomical Journal} can be viewed at
{\tts www.astro.washington.edu/astroj/lemon/yp.html}. Note that these
are {\it submitted} and not necessarily {\it accepted} articles.
The pages come with links to email the authors, which is very helpful if
you wish to ask some of them for a copy.
There is also a page with ``Titles and abstracts of ApJ Letters
accepted but not yet published'' at
{\tts cfa-www.harvard.edu/aas/apjl\_abstracts.shtml}.

A ``Distributed Database of Online Astronomy Preprints and Documents''
is currently in the early stages of development. While not functional as
yet, there is a web page describing the project. People wanting to monitor
progress, or make suggestions, are invited to look at the URL~
{\tts doright.stsci.edu/Epreps}.

\subsection{Preprint Lists from NRAO and STScI} \label{step}

Although the LANL/SISSA server has become very popular recently, there
are still a large number of preprints being distributed {\it only} on paper.
The ideal places to get reasonably complete listings of these are the
following.

The STEPsheet (``Space Telescope Exhibited Preprints'') is a list of all preprints
received during the last two weeks at the STScI and is prepared by its
librarian, Sarah Stevens-Rayburn.  It is delivered by email, and
subscription requests should be sent to {\tts library@stsci.edu}. Each list
contains well over 100 titles. Note that the preprints themselves are
not distributed by the STScI librarian and must be requested from the
individual authors.  The full current STScI database contains everything
received in the last several years, along with all papers received since 1982
and not yet published.  Both databases are searchable at
{\tts sesame.stsci.edu/library.html}.

The RAPsheet (``Radio Astronomy Preprints'') is a listing of all preprints
received in the Charlottesville library of the National Radio Astronomy
Observatory (NRAO) in the preceding two weeks (contact: {\tts library@nrao.edu}).
Interested persons should request copies of preprints from the authors.
The tables of contents of all incoming journals and meeting proceedings are
perused in order to find published references and to update the records.
A database of preprints received since 1986, along with their added references,
and including unpublished ones since 1978, is
also searchable at {\tts libwww.aoc.nrao.edu/aoclib/rapsheet.html}.

\subsection{Electronic Journals}

Most of the major astronomical journals are now available over the
Internet. While they tended to be freely available for a test period
of one to two years, most of them now ask a license which can be obtained
for free only if the host institute of the user is subscribed to the
printed version. Nevertheless, for most journals the tables of contents
are accessible on the WWW for free (see \S\ref{tocs}).
A very useful compilation of links to electronic journals and
other bibliographical services is provided at
{\tts www.ucm.es/info/Astrof/biblio.html}.

Under the ``AAS Electronic Journal Project''
({\tts www.aas.org/Epubs/eapjl/eapjl.html}; cf. \cite{boyce97})
the Astrophysical Journal (ApJ) had been made available electronically
since 1996 ({\tts www.journals.uchicago.edu/ApJ/journal}).
As of April 1997, access to the full text of the ApJ Electronic Edition
is available only to institutional and individual subscribers. What you
{\it can}~ do without subscribing, is to browse the contents pages of even the
latest issues. The policy of AAS is to sell a license for the whole set
of both back and current issues. Licensees will have to keep paying
in order to see any issues of the journal. In return AAS will continue
to maintain electronic links between references, a facility which
the database of scanned back journals at the ADS article service
does not offer.

The electronic {\it Astronomical Journal}
({\tts www.journals.uchicago.edu/AJ/journal}) came on-line in January 1998.
{\it Astronomy \& Astrophysics} (A\&A) and its
{\it Supplement Series} (A\&AS) are available at
{\tts link.springer.de/link/service/journals/00230/tocs.htm}, and
at {\tts www.edpsciences.com/docinfos/AAS/OnlineAAS.html}, respectively.
Abstracts of both journals can be viewed at
{\tts cdsweb.u-strasbg.fr/Abstract.html}.

The journal {\it New Astronomy}, initially designed to appear electronic-only
released its first issue in September 1996
(see {\tts www.elsevier.nl/locate/newast}).
One may subscribe to a service alerting about new articles appearing
in {\it New Astronomy} by sending email to {\tts newast-e@elsevier.nl} with
{\tts Subj:}~{\tts subscribe newast-c}.

Several other journals are available electronically, e.g. the {\it Publications
of the Astronomical Society of the Pacific} (PASP; {\tts
www.journals.uchicago.edu/PASP/journal}); the {\it Proceedings
of the Astronomical Society of Australia} (PASA) has started an experimental web
server, beginning from Vol.~14\,(1997), at {\tts www.atnf.csiro.au/pasa};
Pis'ma Astronomicheskii Zhurnal offers English abstracts 
at {\tts hea.iki.rssi.ru/pazh}, and Bulletin of the American Astronomical 
Society (BAAS) at {\tts www.aas.org/publications/baas/baas.html}.
The IAU ``Informational Bulletin of Variable Stars'' (IBVS) has been
scanned back to its volume 1 (1961) and is available at
{\tts www.konkoly.hu/IBVS/IBVS.html}. One volume (\#\,45) of the 
Bulletin d'Information of CDS (BICDS) is at {\tts cdsweb.u-strasbg.fr/Bull.html}.

The {\it Journal of Astronomical Data} (JAD), announced during the 22$^{nd}$
IAU General Assembly in 1994 as the future journal for the publication of
bulky data sets on CD-ROM, has produced three volumes on CD-ROM since
October 1995, available at cost from {\tts www.twinpress.nl/jad.htm}.

More and more electronic journals are linking their references directly
to the ADS Abstract Service, thus working toward a virtual library
on the Internet.

\subsection{Tables of Contents} \label{tocs}

The ADS abstract service (\S\ref{absserv}) offers to browse specific volumes
of journals via the {\tts TOC QUERY} button on URL
{\tts adswww.harvard.edu/abstract\_service.html}.
For those journals not accessible by the {\tts TOC QUERY} you
can use the {\tts BIBCODE QUERY}, e.g. to browse the
{\it Bull.~Astron.~Soc.~India}, search for
bibcode {\tts 1997BASI...25}.

{\tt UnCover} at the ``Colorado Alliance of Research Libraries'' (CARL)
is a database of tables of contents of over 17,000 multidisciplinary
journals. It can be accessed at URL {\tts uncweb.carl.org}, or via telnet
to {\tts pac.carl.org} or {\tts database.carl.org}. It contains article
titles and authors only, and offers keyword searches for both.
Data ingest started in September 1988.
In late 1997 it included more than 7,000,000 articles, and 4,000 articles
were added daily.  About sixty astronomy-related journals are
present, including ApJ, A\&A, AJ, ApJS, MNRAS, PASP, Nature, Science, and
many others which are less widely distributed.
One of the most useful features is that the contents pages of individual journals
can be viewed by volume and issue. This provides an important
independent resource to monitor the current astronomical literature, especially
if one's library cannot afford to subscribe to more than the major journals.
Copies of all retrieved articles can be ordered (by FAX only)
for a charge indicated by the database. UnCover also provides links
to access other library databases and to browse the
library catalogues of several North American libraries.

For some of the journals not covered by bibliographic services
the Publishers' web pages may offer tables of contents, like e.g.\
for the {\it Monthly Notices of the Royal Astronomical Society} (MNRAS) at
{\tts \verb*Cwww.blackwell-science.com/~cgilib/jnlpage.bin?Journal=MNRASC} \linebreak[4]
{\tts \verb*C&File=MNRAS&Page=contentsC}.
To browse {\it Chinese Astronomy \& Astrophysics} click on
``Contents Services''
at {\tts www.elsevier.com/inca/publications/store/5/8/5/}.
For {\it Astrophysics \& Space Science} see
{\tts www.wkap.nl/jrnltoc.htm/0004-640X}.
The National Academy of Sciences of the USA offers the contents
pages and full texts of its Proceedings and colloquia
at {\tts www.pnas.org}. The American Physical Society has its
{\it Reviews of Modern Physics} at {\tts rmp.aps.org}.
The {\it Icarus} journal offers its tables of contents and lists of
submitted papers at the URL {\tts astrosun.tn.cornell.edu/Icarus/}.
Tabular and other data
from papers in Icarus are published on the AAS CD-ROMs.

\subsection{``Grey Literature'': Newsletters, Observatory Publications, etc.} \label{greylit}

A compilation of links to various astronomical newsletters is provided
by P.~Eenens at \linebreak[4]
{\tts \verb*Cwww.astro.ugto.mx/~eenens/hot/othernews.htmlC}.
In 1994/95, Cathy Van Atta at NOAO (now retired) and a few others
prepared a list of astronomical Newsletters. S.~Stevens-Rayburn (STScI)
has put this list on URL~ {\tts sesame.stsci.edu/lib/NEWSLETTER.htm}
and invites volunteers to complete and update the information.
A catalogue of over 4,000 individual Observatory Publications,
ranging from the 18th century to the present, has been
prepared by Brenda Corbin and is available from within the USNO
Library Online Catalog ``Urania'' (click on ``Library Resources'' at
~{\tts www.usno.navy.mil/library/lib.html}).

A list of IAU Colloquia prepared by STScI librarian S.~Stevens-Rayburn
is offered at \linebreak[4]
{\tts sesame.stsci.edu/lib/other.html}.
A Union List of Astronomy Serials II (ULAS II) compiled by Judy L.~Bausch
can be searched at {\tts sesame.stsci.edu/lib/union.html}.
It provides information on $\sim$2300 (primarily) non-commercial
publications of observatories and institutions concerned with
research in astronomy. For each item, it lists the holding records
of 42 contributing libraries, representing the most comprehensive
astronomical collections in North America, with selected holdings from
China, Europe, India, and South America as well.

A database of book reviews in astronomy from 1987 to the present
has been prepared by Marlene Cummins and is available at
{\tts www.astro.utoronto.ca/reviews1.html}.

\subsection{Library Holdings}

The card catalogues of hundreds of libraries from all disciplines (including
the Library of Congress, which can be accessed on the WWW at
{\tts lcweb2.loc.gov/catalog}) are available over the net, and the number
is continuously growing. The Libweb directory at
{\tts sunsite.berkeley.edu/Libweb} lists library catalogues which are
accessible on the WWW.  \linebreak[4] Libweb is frequently updated and
currently provides addresses of over 1700 libraries from 70 countries.
A more complete listing of astronomy-related libraries can be found
at \linebreak[4] {\tts www.stsci.edu/astroweb/cat-library.html},
the ``Libraries Resources'' section of the AstroWeb.
The STScI library holdings are searchable at {\tts stlibrary.stsci.edu/html}, 
and those of ESO at {\tts www.eso.org/libraries/webcat.html}.

\section{Directories and Yellow-Page Services}  \label{directyellow}

Occasionally, if not frequently, one needs to search for the e-mail of an
astronomer somewhere
in the world. There are many ways to find out and several may have to be tried.
You should start with the ``RGO email directory''.
It is maintained by C.R.~Benn and R.~Martin and is made up of three parts.
One is a list of $\sim$13,000 personal emails, another one offers
phone/FAX numbers and emails or URLs of $\sim$950 astronomical research
institutes, and a third one has postal addresses for $\sim$650
institutions. You should make sure that your departmental secretary knows about
the latter two!  All lists are updated frequently and available for
consultation at {\tts www.ast.cam.ac.uk/astrosearch.html}.  The
impatient and frequent user of these directories should draw a local
copy from time to time (three files at
{\tts ftp.ast.cam.ac.uk/guide/astro*.lis}), or
ask your system manager to install a site-wide command to interrogate
these lists and draw updated copies from time to time.
To request inclusion in this directory or communicate updated addresses,
send a message to {\tt email@ast.cam.ac.uk}.

Be sure that you never (ab)use such lists of thousands of addresses
to send your announcements to the entire list. You will most likely
offend the majority of the recipients who are not interested in your
message, which may be even regarded as ``spamming'' (see e.g.
{\tts \verb*Ccdsweb.u-strasbg.fr/~heck/spams.htmC} for a collection of links to
defend users from unsolicited email).
However, these email directories may be useful for the legitimate task of
selecting a well-defined subset of researchers as a distribution list
for specific announcements.

The RGO email guide depends on personal and institutional input for its
updates and turns out to be fairly incomplete especially for North American
astronomers. Complete addresses for AAS members can be found in the AAS
Membership Directory which appears in print annually and is distributed
to AAS members only.
The AAS membership directory has been put on-line experimentally and made
searchable at {\tts directory.aas.org}, but it cannot be downloaded entirely.

Since 1995 the IAU membership directory has been accessible at the LSW Heidelberg 
web site, but in early 1998 it moved to {\tts www.iau.org/members.html}.
The address database is managed by the IAU office in Paris
({\tts www.iau.org}), and requests for updates have to be sent there
by email ({\tts iau@iap.fr}). From these the IAU office
prepares an updated database every few months which is then put on-line.
Hopefully this web address for the IAU membership directory will remain
stable in the future, and not change every three years with the election
of a new General Secretary at each IAU General Assembly.

The European Astronomical Society (EAS) is preparing its membership directory
under URL {\tts www.iap.fr/eas/directory.html}. It provides links
to membership directories of several national astronomical
societies in Europe (see {\tts www.iap.fr/eas/societies.html}).

The ``Star*sFamily'' of directories is maintained at CDS (\cite{heck97})
and divided into three parts.
``StarWorlds'' ({\tts \verb*Ccdsweb.u-strasbg.fr/~heck/sfworlds.htmC})
offers addresses and many practical details for $\sim$6,000 organizations,
institutions, associations, companies related to astronomy and space sciences
from about 100 countries, including about 5,000 direct links to their homepages.
``StarHeads'' ({\tts \verb*Ccdsweb.u-strasbg.fr/~heck/sfheads.htmC})
is a compilation of links to personal WWW homepages of about
4500 astronomers and related scientists.  For ``StarBits'' see \S\ref{Dic}.

Another way to search for email addresses is a search engine 
(mirrored at various sites) which can be accessed via 
{\tts telnet bruno.cs.colorado.edu}, login as user {\tts netfind}.
You should give either first, last or login name of the person you
look for, plus keywords containing the institution and/or the 
city, state, or country where the person works.

There are more, rather ``informal'' ways of tracing emails of
astronomers. One is to check whether they have contributed preprints to
the SISSA/LANL server (\S\ref{ppserv}) recently. If so, the address from
which they sent it will be listed in the search result returned to you.
Another way is the command~ {\tts finger xx@node.domain}, where {\tts xx}
is either the family name or a best guess of a login name of the
individual you seek. However, some nodes prefer ``privacy'' and disable
this command. One last resort is to send an email to {\tts postmaster}
at the node where you believe the person is or used to be.

\section{Meetings and Jobs}   \label{meetjob}

Since about 1990, the librarian at the Canada-France-Hawaii
Telescope (CFHT, Hawaii), E.~Bryson, has maintained a list of
forthcoming astronomical meetings, including those back to Sept.~1996,
at the URL {\tts cadcwww.dao.nrc.ca/meetings/meetings.html}.
One may subscribe to receive updates of this list of meetings
by request to \linebreak[4]  {\tts library@cfht.hawaii.edu}.
It is now the most complete reference in the world for future astronomy meetings.
Organizers of meetings should send their announcements
to {\tts library@cfht.hawaii.edu} to guarantee immediate and world-wide
diffusion. Official meetings of the IAU and some other meetings of
interest to astronomers are announced in the IAU Information Bulletin
(see {\tts www.iau.org/bulletin.html}). A list of all past IAU
Symposia is available at {\tts www.iau.org/pastsym.html}.

Probably the most complete collection of job advertisements in
astronomy is the ``AAS Job Register'' at
{\tts www.aas.org/JobRegister/aasjobs.html}.
The European Astronomical Society (EAS) maintains a Job Register
at {\tts www.iap.fr/eas/jobs.html}.
STARJOBS is an electronic notice board maintained at the Rutherford Appleton
Laboratory. The service is co-sponsored by the EAS
and includes announcements of all European astronomical jobs
notified to the Starlink astronomical computing project
({\tts telnet} to {\tts star.rl.ac.uk}, login as {\tts starjobs}).
One can also copy (via {\tt ftp}) the complete list of jobs as the file
{\tts starlink-ftp.rl.ac.uk:/pub/news/star\_jobs}.
For employment opportunities in the Space Industry consult
the URL~ {\tts www.spacejobs.com}.
The AstroWeb offers links to job offers at
{\tts www.cv.nrao.edu/fits/www/yp\_jobs.html}.

\section{Dictionaries and Thesauri} \label{Dic}

The {\it Second Dictionary of the Nomenclature of Celestial Objects} (see
Lortet, Borde \& Ochsenbein (1994)) 
can be consulted at {\tts vizier.u-strasbg.fr/cgi-bin/Dic} or
via telnet to {\tts simbad.u-strasbg.fr} (login as {\tts info}, give
{\tts <CR>} as password, then issue the command {\tts info cati XXX} to inquire
about the acronym {\tts XXX}).
Authors of survey-type source lists are strongly encouraged to check that
designations of their objects do not clash with previous namings and are
commensurate with IAU recommendations on nomenclature
{\tts cdsweb.u-strasbg.fr/iau-spec.html}. In order to guarantee that
designations of an ongoing survey will not clash with other names,
authors or PIs of such surveys should consider to pre-register an acronym
for their survey some time before publication at the URL given above.
Information on over 4000 acronyms is provided.

Independently, a list of ``Astronomical Catalog Designations'' has been
prepared by INSPEC (see {\tts www.iee.org.uk/publish/inspec/astro\_ob.html}).
It is less complete than the CDS version and deviates in places
from the IAU recommendations.

The dictionary ``StarBits'', maintained by A.~Heck (Strasbourg), offers
$\sim$120,000 abbreviations, acronyms, contractions and symbols from
astronomy and space sciences and related fields. It is accessible
at {\tts \verb*Ccdsweb.u-strasbg.fr/~heck/sfbits.htmC}.
Astronomers are invited to consult this dictionary to avoid assigning an
acronym that has been used previously.

On behalf of the IAU several librarians of large astronomical institutions
prepared ``The Astronomy Thesaurus'' of astronomical terms (\cite{shobb93},
and later its ``Multi-Lingual Supplement'' (\cite{multi95})
in five different languages (English, French, German, Italian, and Spanish).
It is freely available at {\tts www.aao.gov.au/library/thesaurus}.
It may be useful in many respects, e.g., to translate astronomical
terms, to aid authors in better selection of keywords for their papers,
and to help librarians improve the classification of publications.
The Thesaurus is available via anonymous {\tt ftp} for DOS, MAC and Unix systems
({\tts www.aao.gov.au/lib/thesaurus.html}); it has not been
updated for some years, but M.\,Cummins
({\tts astlibr@astro.utoronto.ca}) is currently in charge of it
and appreciates comments about its future.

As an aside I mention the ``Electronic Dictionary of Space
Sciences'' by J.~Kleczek and H.~Kleczkov\'a, who have collected
several 10,000 of words, synonyms and expressions from astronomy,
space sciences, space technology, earth- and atmospheric sciences
and related mathematics, physics, and engineering fields in 
five languages\.: English, French, German, Spanish, and Portuguese 
(see {\tts www.twinpress.nl/edss.htm} for an electronic version at cost).
The ``Oxford English Dictionary''
(OED; {\tts www1.oed.com/proto/}) is currently being revised to
include a far more comprehensive set of astronomical terms than
before (see the OED Newsletter at~ {\tts www1.oup.co.uk/reference/};
\cite{mahoney98}).

\section{Miscellaneous}

\subsection{Astronomical Software} \label{softw}

For an introduction to publicly available astronomical software and
numerical libraries see \cite{feigmurt92}.
The ``Astronomical Software and Documentation Service'' (ASDS)
at {\tts asds.stsci.edu/asds}  contains links to the major
astronomical software packages and documentation. It allows one to
search for keywords in all the documentation files available.
The {\it Statistical Consulting Center for Astronomy} at Penn State University
({\tts www.stat.psu.edu/scca/homepage.html})
offers advice and answers to frequently asked
questions about statistical applications. The ``StatCodes'' (of
statistical software for astronomy and related fields, at
{\tts www.astro.psu.edu/statcodes}) are now also in ASDS.

\subsection{Observatory and Telescope Manuals} \label{obsmans}

The librarian at CFHT (Hawaii), E.~Bryson, has collected observatory
and telescope manuals in electronic form from all over the world.
Several dozen such documents are now available through the
ASDS (\S\ref{softw}) at {\tts asds.stsci.edu/asds}.
The documents may be searched by keywords.

\subsection{IAU Circulars, Minor Planets, ATEL, and Ephemerides}  \label{iaucirc}

Information about time-critical phenomena has been distributed
by the IAU ``Central Bureau for Astronomical Telegrams'' (CBAT;
{\tts cfa-www.harvard.edu/cfa/ps/cbat.html}) 
via ``IAU Circulars'' since October 1922.
The IAU ``Minor Planet Center'' (MPC;
{\tts cfa-www.harvard.edu/cfa/ps/mpc.html}) is responsible for the
collection and dissemination of astrometric observations and orbits
for minor planets and comets, via the ``Minor Planet Electronic
Circulars'' (MPEC; {\tts cfa-www.harvard.edu/cfa/ps/services/MPEC.html}),
distributed on paper as the ``Minor Planet Circulars'' since before 1947.
Through collaboration with CBAT and MPC, the ADS Abstract Service
includes electronic circulars of these two sites.
The title, author, and object names are freely available through
the ADS, and the whole text of each circular is indexed by ADS so that
it is found on searches.
The references returned from the ADS include an on-line link to the full
circular at CBAT and MPC.

In December 1997 the ``Astronomer's Telegram'' (ATEL; \cite{1998PASP..110..754R})
was released. It is a web-based publication system for short notices on
time-critical information and is available at {\tts fire.berkeley.edu:8080/}).
Submission of telegrams is restricted to professional astronomers and requires 
special permission. The potential authors are given a special authentication
code to be used at the time of submission, which is entirely automated and
unmoderated, without any human intervention. Readers can freely access the telegrams
or ask to be on a mailing list to receive telegrams within minutes of
their submission. During the first six months of its existence about five telegrams
per month were received.

Ephemerides and orbital elements of comets and minor planets can be
consulted e.g.\ at {\tts cfa-www.harvard.edu/iau/Ephemerides/index.html}
or at JPL's HORIZONS system ({\tts ssd.jpl.nasa.gov/horizons.html}).
A free, interactive program for fancy calculations of ephemeris,
visibility curves from any site on Earth, and graphical displays of
finding charts based on the HST Guide Star Catalogue,
orbits of solar system bodies, views of Earth and Moon, and much more
is available (for X-Windows systems with Motif) for download at
{\tts \verb*Ciraf.noao.edu/~ecdowney/xephem.htmlC}.

The ``Astronomy Calculator'', available at 
{\tts \verb*Cw3.one.net/~rback/frames.htmlC},
aspires to provide general information about the
phases of the moon, lunar eclipse, next annual meteor shower and planets.

\subsection{Atomic Data}   \label{atomic}

A multitude of links to databases containing atomic and molecular data
can be found at
{\tts cfa-www.harvard.edu}, ~in~ {\tts .../amp/data/otherdb.html} ~or~
{\tts \verb*C.../~esmond/amdata.htmlC}.

The ``Opacity Project'' (OP) at {\tts vizier.u-strasbg.fr/OP.html} offers
extensive atomic data required to estimate stellar envelope opacities.

The original implementation of the ``Vienna Atomic Line Data Base'' (VALD;
Piskunov et al.~(1995) is available at
{\tts \verb*Cwww.ast.univie.ac.at/~weiss/vald.htmlC}.
The VALD manual (at {\tts plasma-gate.weizmann.ac.il/VALD.html})
is part of a summary of ``Databases \linebreak[4] for Atomic and Plasma Physics'' (DBfAPP)
of the Weizmann Institute (see \linebreak[4] {\tts plasma-gate.weizmann.ac.il/DBfAPP.html}).
An updated and improved interface for VALD is under construction (at
{\tts www.astro.uu.se/vald}), though actual data traffic will continue
to be handled via e-mail.
This page will include all necessary links including documentation,
registration form, requests forms and examples.
Send inquiries to F.~Kupka (email: {\tts valdadm@jan.ast.univie.ac.at}).

\subsection{Libraries}

Two distribution lists are available specifically for astronomy librarians
to share relevant information on widely varying subjects.
Astrolib (started in 1988) with $\sim$200 members is managed
by E.\,Bouton ({\tts library@nrao.edu}), librarian at the National Radio Astronomy
Observatory (NRAO).  The European Group of Astronomy Librarians (EGAL), is
managed by I.\,Howard ({\tts howard@ast.cam.ac.uk}), librarian at the
Royal Greenwich Observatory (RGO).
J.\,Regan ({\tts library@mso.anu.edu.au}) is currently trying to set up
a group in Asia and the Pacific Rim.
People wishing to post announcements to libraries including physics and
mathematics departments should get in touch with the moderator of the
email distribution list of PAMnet, a network of Physics, Astronomy and
Mathematics librarians. Send your inquiry to {\tts david.e.stern@yale.edu}.

U.\,Grothkopf, ESO librarian, maintains a list of
names, addresses, phone and fax numbers, email address and homepage URLs
of astronomy librarians and libraries world-wide. A useful search engine
({\tts www.eso.org/libraries/astro-addresses.html}) can find librarians
even from incomplete information.
Librarians not on the list are encouraged to send information to
{\tts esolib@eso.org}.

\subsection{Astronomy Education on the Internet}  \label{education}

Given that the Internet offers almost unlimited possibilities for
interactive courses, more and more of these can be found on the web,
see e.g.\ \cite{benacc98}.  The ``AstroEd'' page, at
{\tts www-hpcc.astro.washington.edu/scied/astro/astroindex.html},
provides links to on-line astronomy education resources and
some on-line courses. The National Research Council's project 
``Resources for Involving Scientists in Education'' (RISE)
features a web site at~ {\tts www.nas.edu/rise/examp.htm}.
An interesting example is given by the University of Oregon 
({\tts www.zebu.uoregon.edu}), where an astronomy book 
is being developed in ``hypertext''. A further advantage of these 
``books'' is their potential of being kept up-to-date by a 
groups of professionals.

The European Southern Observatory (ESO) organized an educational
programme called ``Astronomy On-line'' in December 1997. Its web
pages are still available at {\tts www.eso.org/astronomyonline/}
(\cite{albrecht98}).

\subsection{Others}  \label{others}

D.~Verner's compilation of people mentioned in acknowledgments of papers
in major astronomical journals can be accessed from
{\tts \verb*Cwww.pa.uky.edu/~verner/aai.htmlC}.

\section{Issues for the Future} \label{future}

The Internet has been with us for only about a decade. Users of the
WWW should be aware that there is still more information, literature,
data, etc., existing only in printed form, than is available
on the Internet.
While the possibilities of information and data retrieval have
advanced at a tremendous pace in recent years,
there is an infinite number of possible improvements.
I shall mention only a few very subjective ones as an example here.

The increasing presence of commercial companies on the Internet is
both an enrichment and a plague, the latter because more and more frequently
unsolicited emails are being sent to global distribution lists with
commercial offers. While this is annoying, and measures should be
taken against it (see {\tts \verb*Ccdsweb.u-strasbg.fr/~heck/spams.htmC}),
I do not think that it is a reason for astronomers to refrain from being
listed in email guides or from making these guides available among colleagues.
The damage to easy communication among scientists would be too severe.

The transition from printing large tables on paper to publishing them in
electronic form
(\S\ref{catal}) either in the ADC and CDS archives (or on the AAS CD-ROMs),
raises the question about the future of marking-up tables for
printing. For many years authors have been obliged to convert their data
tables to \LaTeX\ format. Ironically, AAS requests a charge of US\$ 50 for
the service to convert the data tables back to plain ASCII format for
publication on their CD-ROM, except for tables marked up with the AASTeX
macros (see {\tts ftp://ftp.aas.org/cdrom/guidelines.html}).
It may be anticipated that the ``publication'' of
tables in electronic form will eventually release authors from
this task. However, special non-ASCII symbols, like e.g.\ Greek letters,
will require to be transliterated to ASCII characters in the electronic version.

Unfortunately the journals in astronomy do not yet oblige authors
to provide their tabular data to a data centre, as a requisite for
publication. An agreement between all major journals and the data centres
ADC and CDS is highly desirable, not only for the sake of the completeness
of their electronic archives of tabular and catalogue data, but also
to remedy the following problem.
The clearing house of the IAU Task Group on Astronomical Designations of
IAU Commission 5 ({\tts cdsweb.u-strasbg.fr/iau-spec.html}) has frequently
come across unconventional namings of astronomical objects causing
confusion and redundancy of names in object databases like NED and SIMBAD.
My experience in the Task Group was that the standard refereeing system
of journals does not help to avoid this problem.
Ideally, these tables should be run through an
automatic cross-checking routine prior to publication or acceptance.
For this purpose they should have at least two identifiers (a name
and a coordinate) and could then be compared with databases like SIMBAD, NED or
LEDA, in order to check the consistency of names and coordinates, and
perhaps even part of the data. Of course this is useful only if the objects
were known previously.

\section{Concluding Remarks}

The Internet and World Wide Web have added just another medium
for fast access to large amounts of information. It can save
researchers lots of time in retrieving the required information
and allows access to unexplored data which are worth many
research projects in their own right.
However, the flood of information
on the web has become so large that now, when searching for a given
piece of information, we are about to spend more time in browsing the web
than we used to need searching in the library a decade ago
(when the amount of available information was substantially less).

In the early years of networking we were happy when we could
get electronic copies of astronomical catalogues without the
delays through shipments of tapes. Now we are so flooded with
them that in the rush of using many of them at the same time
we sometimes forget that each one of them is telling us
a different story. We must still read their detailed documentation
if we want to derive reliable results from the available data.
We have gone a good part of the way already to the point where all past
issues of the major astronomical journals will be available electronically on
the web. However, network saturation still keeps us from skimming a journal
in the way that we could in the library.

A compromise has yet to be found between a rigorous
refereeing system of web pages (as proposed by some) and the absolute
liberty we currently ``enjoy'' in offering our own information and
expressing our interests on the web.
Many people have tried in recent years to offer guides to certain
parts of astrophysical information, and the present article is just
another example. The challenge for the future is how to protect
ourselves from too much redundant or superseded information. While preparing
this paper I came across many web pages which at first sight looked
promisingly complete. However, when the last update (if given at all)
was more than about a year ago, I usually refrained from quoting it here,
because of the danger that it would not be maintained any more,
or that it would offer too many outdated links.
Perhaps a step towards reducing this danger could
be a web browser that automatically recognized the date of latest update
of a web document and would allow to set filters on that date in
a search for relevant links. Actually, the AltaVista search engine
({\tts www.altavista.digital.com}) allows a range of ``last modified''
dates to be used in advanced searches.

\begin{acknowledgments}
I am grateful to the School organizers for the opportunity
to give these lectures and for their financial support.
My special thanks go to Clive Davenhall, Elizabeth Griffin, and 
Andr\'e Fletcher for their careful reading of the manuscript.
Useful information or comments on the text were provided by
P.~Boyce, P.~Eenens, G.~Eichhorn, D.~Fullagar, G.~Giovannini,
D.~Golombek, C.\,S.~Grant, U.~Grothkopf, R.J.~Hanisch, M.~Irwin, 
F.~Kupka, S.~Kurtz, N.~Loiseau, A.~Macdonald, F.~Murtagh, F.~Ochsenbein,
G.~Paturel, M.~Schmitz, S.\,A.~Trushkin, M.~Tsvetkov, and M.\,J.~West.
My apologies to those I forgot to mention here.
\end{acknowledgments}

\end{document}